\documentclass[ prx, twocolumn, nofootinbib, floatfix]{revtex4-2} 

\usepackage{times}
\usepackage{graphicx}
\usepackage{amsmath}
\usepackage{enumitem}
\usepackage{amssymb}
\usepackage{array}
\newcolumntype{L}[1]{>{\raggedright\let\newline\\\arraybackslash\hspace{0pt}}m{#1}}
\newcolumntype{C}[1]{>{\centering\let\newline\\\arraybackslash\hspace{0pt}}m{#1}}
\newcolumntype{R}[1]{>{\raggedleft\let\newline\\\arraybackslash\hspace{0pt}}m{#1}}
\usepackage{mathrsfs}
\usepackage{dsfont}
\usepackage{appendix}
\usepackage{multirow}
\usepackage{dsfont}
\usepackage{tikz}
\usepackage{amsfonts}
\usepackage{amssymb}
\usepackage{pgflibraryarrows}
\usepackage{pgflibrarysnakes}
\usetikzlibrary{decorations.pathmorphing}
\usepgflibrary{arrows.meta}

\usepackage{subfigure}
\usepackage{bbold}
\newcommand{\D}{\mathrm{d}}

\newcommand{\Omgea}{\Omega}

\usepackage{accents}
\usepackage{braket}

\usepackage{cancel}

\graphicspath{{Figures/}}
\allowdisplaybreaks

\usepackage{xtab,afterpage}
\usepackage{hyperref}
\definecolor{blueblue}{RGB}{21,47,181}

\hypersetup{ linktoc=all,
    colorlinks, linkcolor={blueblue},
    citecolor={blueblue}, urlcolor={blueblue}
}

\begin{document}

\title{Thermality, causality, and the quantum-controlled Unruh-deWitt detector}

\author{Joshua Foo}
\email{joshua.foo@uqconnect.edu.au}
\author{Sho Onoe}
\affiliation{Centre for Quantum Computation and Communication Technology, School of Mathematics and Physics, The University of Queensland, St. Lucia, Queensland, 4072, Australia}
\author{Robert B. Mann}\email{rbmann@uwaterloo.ca}
\affiliation{Department of Physics and Astronomy, University of Waterloo, Waterloo, Ontario, Canada, N2L 3G1}
\affiliation{Perimeter Institute, 31 Caroline St., Waterloo, Ontario, N2L 2Y5, Canada}
\author{Magdalena Zych}
\email{m.zych@uq.edu.au}
\affiliation{Centre for Engineered Quantum Systems, School of Mathematics and Physics, The University of Queensland, St. Lucia, Queensland, 4072, Australia}

\date{\today}

\begin{abstract}
{Particle detector models such as the Unruh-deWitt detector are widely used in relativistic quantum information and field theory to probe the global features of spacetime and quantum fields. These detectors are typically modelled as coupling locally to the field along a classical worldline. In this paper, we utilize a recent framework which enables us to prepare the detector in a quantum-controlled superposition of trajectories, and study its response to the field in finite-temperature Minkowski spacetime and an expanding de Sitter universe. Unlike a detector on a classical path which cannot distinguish these spacetimes, the superposed detector can do so by acquiring nonlocal information about the geometric and causal structure of its environment, demonstrating its capability as a probe of these global properties.}
\end{abstract}
\maketitle

\section{Introduction}
In quantum field theory (QFT), the physical nature of phenomena like particle production and long-range correlations is grounded in an ability to couple local probes to the field, which can subsequently perform measurements of it. The study of measurements in QFT continues to be a fruitful and ongoing research area, linking diverse subjects from ranging from causality \cite{Sorkin:1993gg}, quantum optics and quantum information \cite{martinPhysRev.115.1342,ramonPhysRevD.103.085002,funaiPhysRevD.100.065021}, to curved spacetime settings \cite{Fewster:2018qbm}. 

A simple and well-known approach for enacting measurements of a quantum field is the Unruh-deWitt (UdW) detector. The detector is typically modelled as an idealised two-level system whose internal states couple to a massless scalar field, which approximates the light-matter interaction under the neglect of angular momentum exchange \cite{unruh1984happens,davies2002detection,birrell1984quantum}. In settings which involve arbitrary relativistic trajectories  \cite{satz2007then,louko2006often,lin2007backreaction,sriramkumar1996finite} and curvature \cite{louko2008transition,hodgkinson2012static,ng2014unruh,ng2017over}, the model is particularly useful because it provides operational meaning to the notion of a `particle', through the excitations it experiences via the interaction. Phenomena which exemplify this include the Unruh and Gibbons-Hawking effects. The former predicts that a uniformly accelerated detector in Minkowski spacetime perceives the vacuum state to be thermal at the Unruh temperature:
\begin{align}\label{TU}
    T_U &= \frac{\kappa}{2\pi},
\end{align}
whereas an identical detector traversing an inertial worldline registers no particles \cite{unruh1976notes}, as expected for the vacuum state. There is an analogous situation in the de Sitter universe, where a detector on a geodesic path likewise detects the conformally coupled  vacuum state to be thermal at the same temperature as in Eq.~\eqref{TU}, where $\kappa$ now quantifies the expansion rate of the universe \cite{gibbons1977cosmological}. Hence, according to a detector traversing a classical worldline and fully characterised by its response to quantum fields, flat and de Sitter spacetimes in the above scenarios are operationally equivalent i.e.~both give rise to a thermal bath at the finite temperature,
\begin{align}
    T_T &= \frac{\kappa}{2\pi}.
\end{align}
Contrary to the intuition that acceleration fully determines the thermal response of the detector, it has been recently demonstrated that a UdW detector travelling in a superposition of accelerated trajectories in general does \textit{not} yield a thermal response \cite{fooPhysRevD.102.085013,barbadoPhysRevD.102.045002}. In particular, even if the individual trajectories have the same proper acceleration and therefore each of them would  yield the same thermal state of the detector --  their superposition is sensitive to nonlocal field correlations between the trajectories, which notably, depend on the causal relations between them. These correlations can perturb the final detector state away from thermalization.  

In this article, we apply the superposed detector model to the above scenarios: a thermal field state in Minkowski spacetime and the conformally coupled vacuum state in an expanding de Sitter spacetime. Our results show that a single UdW detector in a quantum superposition of trajectories can differentiate between these two spacetimes, which is impossible for a single detector on a classical worldline \cite{rabochaya2016quantum,garbrecht2004unruh,acquaviva2012unruh,tian2013geometric,singh2013quantum}, and even for two detectors harvesting entanglement from the field, in certain regimes \cite{ver2009entangling,nambu2013entanglement,kukita2017entanglement,tian2014dynamics,tian2016detecting,salton2015acceleration}. We show in particular, that the response is sensitive to the causal relationship between the paths in superposition, signatory of the geometric structure of the background spacetime itself. Due to the different spacetime geometries under consideration as well as the thermalisation processes investigated in this work, our results also show that the quantum-controlled UdW detector model represents an accessible approach for probing the geometric and causal features of spacetime, and connects the research in curved spacetime QFT with quantum information \cite{mann2012relativistic}, quantum control of quantum channels \cite{oi2003interference,chiribella2019quantum}, and quantum thermodynamics \cite{kosloffe15062100,andersdoi:10.1080/00107514.2016.1201896}. 

Our article is arranged as follows: in Sec.\ \ref{sec:II}, we introduce the quantum-controlled detector model first devised in \cite{fooPhysRevD.102.085013,barbadoPhysRevD.102.045002}. In Sec.\ \ref{sec:III} we introduce the field-theoretic details needed for the spacetimes of interest. In Sec.\ \ref{sec:IV}, we study the transition probability of the superposed detector in the respective scenarios. In Sec.\ \ref{sec:V}, we analyse the transition rate of the detector in these spacetimes, before offering some conclusions in Sec.\ \ref{sec:VI}. Throughout this article, we utilize natural units, $\hslash = c = k_B = G =1$.

\section{Detectors in superposition}\label{sec:II}
In this paper, we employ the simplest formulation of the UdW detector model, which is a point-like, two-level system initially prepared in its ground state $|g\rangle$ and interacting with a real, massless scalar field $\hat{\Phi}(\mathsf{x}(\tau))$ pulled back to the worldline $\mathsf{x}(\tau)$ and initially in the state $|\psi\rangle$. Following \cite{fooPhysRevD.102.085013,barbadoPhysRevD.102.045002}, we initialise the detector in an arbitrary superposition of trajectories by introducing a control degree of freedom, whose orthonormal states $|i\rangle_C$ designate the individual paths that the detector takes. The initial state of the combined system is thus
\begin{align}
    |\Psi\rangle_S &= |\phi\rangle \otimes | g \rangle \otimes |\psi\rangle,  
\end{align}
where the control is prepared in the equal superposition of $N$ paths,
\begin{align}
    |\phi\rangle &= \frac{1}{\sqrt{N}} \sum_{i=1}^N | i \rangle_C .
\end{align}
When writing down the initial control superposition state, the tacit assumption is that any phases acquired by the system during the  preparation of the superposition have been absorbed into the basis states $|i \rangle_C$.

The Hamiltonian governing the interaction is given by
\begin{align}
    \hat{H}_\text{int.} (\tau) &= \lambda\sigma(\tau) \sum_{i=1}^N \eta_i(\tau)  \hat{\Phi}(\mathsf{x}_i(\tau)) 
    \otimes |i\rangle\langle i |_C  
\end{align}
where $\lambda\ll 1$ is a weak coupling constant, $\eta_i(\tau)$ is a time-dependent switching function that governs the interaction, $\sigma(\tau) = \sigma^+e^{i\Omega\tau} + \text{h.c}$ is the interaction picture Pauli operator (where $\sigma^+ = |e\rangle\langle g|$) for the detector with energy gap $\Omgea$ between the energy eigenstates $|g\rangle, |e\rangle$ and $\mathsf{x}_i(\tau)$ is the worldline of the $i$th path of the superposition. To leading order in $\lambda$, the state of the detector after evolving from the initial time $\tau_0$ to the final time $\tau_F$, and conditioned upon the control being measured in the state $|\phi \rangle$ (chosen for simplicity but without loss of generality), is given by
\begin{align}
    \hat{\rho}_D &= \begin{pmatrix} 1 - \mathcal{P}_D & 0 \\ 0 & \mathcal{P}_D 
    \end{pmatrix} + \mathcal{O}(\lambda^4)
\end{align}
where the transition probability $\mathcal{P}_D$ is given by 
\begin{align}\label{probability}
    \mathcal{P}_D &= \sum_{i=1}^N \mathcal{P}_{ii,D} + \sum_{i\neq j }^N \mathcal{P}_{ij,D}.
\end{align}
We have expressed the transition probability as a sum of two contributions, given respectively by
\begin{align}
    \mathcal{P}_{ii,D} &= \frac{\lambda^2}{N^2}  \int_{\tau_0}^{\tau_F} \D \tau \int_{\tau_0}^{\tau_F} \D \tau' \chi(\tau ) \overline{\chi}(\tau') \mathcal{W}(\mathsf{x}_i(\tau), \mathsf{x}_i(\tau')) \\
    \mathcal{P}_{ij,D} &= \frac{\lambda^2}{N^2} \int_{\tau_0 }^{\tau_F } \D \tau_i \int_{\tau_0 }^{\tau_F } \D \tau_j' \chi(\tau_i) \overline{\chi}(\tau_j' ) \mathcal{W}(\mathsf{x}_i(\tau_i), \mathsf{x}_j(\tau_j'))
\end{align}
where we have defined $\chi_i(\tau) = \eta_i(\tau) e^{-i\Omega\tau}$ and 
\begin{align}
    \mathcal{W}^{ij}(\mathsf{x}_i(\tau_i) , \mathsf{x}_j(\tau_j')) = \langle \psi | \hat{\Phi}(\mathsf{x}_i(\tau_i) ) \hat{\Phi}(\mathsf{x}_j(\tau_j')) |\psi\rangle
\end{align}
are two-point correlation (Wightman) functions pulled back to the  trajectories $\mathsf{x}_i(\tau_i), \mathsf{x}_j(\tau_j')$ \cite{birrell1984quantum}. Importantly, Eq.\ (\ref{probability}) contains Wightman functions evaluated locally along the individual trajectories ($i = j$) and \textit{non-locally}, between each respective pair of trajectories ($i \neq j$). The $\mathcal{P}_{ij,D}$ terms in the transition probability are equal to the nonlocal correlations between two detectors, each locally coupling to a quantum field along individual classical paths $\mathsf{x}_i(\tau_i)$ and $\mathsf{x}_j(\tau_j')$. Specifically, when one introduces a second detector the reduced bipartite density matrix to leading order in perturbation theory is given by 
\begin{align}\label{reduceddensity}
    \hat{\rho}_{AB} &= \begin{pmatrix} 1 - \mathcal{P}_A -\mathcal{P}_B & 0 & 0 & \mathcal{M} \\ 0 & \mathcal{P}_B & \mathcal{L}_{AB} & 0 \\ 0 & \mathcal{L}_{AB}^\star  & \mathcal{P}_A & 0 \\ \mathcal{M}^\star & 0 & 0 & 0  \end{pmatrix} + \mathcal{O}(\lambda^4).
\end{align}
Specifically, the $\mathcal{P}_{ij,D}$ terms for a single superposed detector are equal to the $\mathcal{L}_{AB}$ terms in the two-detector scenario, these terms quantifying in part the nonlocal field correlations along the respective detector worldlines.

The other quantity of interest is the transition rate of the detector, defined as a derivative of the transition probability of the detector, Eq.\ \eqref{probability}, with respect to the proper time $\tau_F$. Physical interpretation of this quantity is the difference between transition probabilities between two ensembles of identically prepared detectors traversing fixed superposition of trajectories and measured at the proper times $\tau_F$ and $\tau_F+\delta\tau$ in the limit $\delta \tau \to 0^+$ \cite{louko2008transition,foo2020unruhdewitt}. To compute the transition rate, we thus take the switching functions to be Gaussian, $\eta(\tau ) = \exp ( - \tau^2/2\sigma^2)$, in the infinite interaction-time limit ($\sigma \to \infty$), taking $\tau_0 \to -\infty$ and evaluate the derivative of Eq.\ (\ref{probability}) with respect to $\tau_F$, which yields the following expression,
\begin{align}\label{eq:7}
    \dot{\mathcal{P}}_D &=  \frac{2\lambda^2}{N^2} \sum_{i=1}^N \text{Re} \int_0^\infty \D s \:e^{-i\Omega s} \mathcal{W}^{ii}(s)   \nonumber \\
    & + \frac{2\lambda^2}{N^2} \sum_{i\neq j }^N \text{Re}  \int_0^\infty \D s \:e^{-i\Omega s} (\mathcal{W}^{ij} (\tau , \tau - s) + \text{H.c})  . \vphantom{
     \sum{i\neq j}^N}
\end{align}
where $s = \tau-\tau'$. In this paper, we only consider scenarios where the proper times of the paths in superposition are equal; hence the simplification to a common proper time coordinate in Eq.\ (\ref{eq:7}).

\section{Wightman functions for thermal fields, de Sitter spacetime
and accelerated trajectories}
\label{sec:III}
Equations (\ref{probability}) and (\ref{eq:7}) characterise the detector's response to the background scalar field as it traverses different regions of spacetime. For our analysis, we require Wightman functions pulled back to the individual trajectories of the superposition, as well as nonlocal Wightman functions between each respective pair of trajectories. For the former, these possess an identical form in all cases considered:
\begin{align}
    \mathcal{W}_D(s) &= - \frac{\kappa^2}{16\pi^2} \frac{1}{\sinh^2(\kappa s /2 -i\varepsilon)}.
\end{align}
Here, the meaning of $\kappa$ depends on the context (e.g.\ the temperature of thermal state of a field in Minkowski spacetime, or the expansion rate of de Sitter spacetime), while $\varepsilon$ is an infinitesimal regularisation constant. 

In Minkowski spacetime for a thermal field state with temperature  $T_T = \kappa(2\pi)^{-1}$, the nonlocal Wightman functions evaluated between two trajectories separated by the constant distance $L$ are given by \cite{weldon2000thermal}
\begin{align}\label{thermalwightman}
    \mathcal{W}_T(s) &= \frac{\kappa}{16\pi^2L} \bigg[  \coth \left( \frac{\kappa}{2} (L-s + i \epsilon) \right) \nonumber \\
    & + \coth \left( \frac{\kappa}{2} (L+s - i\epsilon) \right) \bigg] .
\end{align}
In the de Sitter universe, we parametrise the detector worldlines with flat slicing coordinates \cite{griffiths2009exact}, yielding the Wightman function
\begin{align}
    \mathcal{W}_\text{dS}(p,s) &= \frac{(\kappa/4\pi)^2}{\exp(\kappa p)(\kappa L/2)^2 - \sinh^2(\kappa s/2 -i \varepsilon)}
\end{align}
which are evaluated in the conformally coupled vacuum \cite{birrell1984quantum} and $p = \tau + \tau'$. The trajectories are separated by the constant co-moving distance $L$. For two accelerated trajectories in parallel motion with proper acceleration $\kappa $ and separated by the constant distance $L$ (as measured by inertial observers), we have \cite{salton2015acceleration}
\begin{align}
    \mathcal{W}_P(p,s) &= \frac{\kappa^2}{16\pi^2} \bigg[ \frac{L\kappa}{2} + i \varepsilon - e^{-p\kappa/2} \sinh( \kappa s/2 ) \bigg]^{-1} \nonumber  \\
    & \times \bigg[ \frac{L\kappa}{2} - i\varepsilon + e^{p \kappa/2} \sinh(\kappa s/2  ) \bigg]^{-1}.
\end{align}

\subsection{Thermality in QFT}
In QFT, thermal states are those  satisfying the Kubo-Martin-Schwinger (KMS) condition \cite{Kubo:1957mj,martinPhysRev.115.1342}. This condition provides a general definition for a thermal state in scenarios where the usual Gibbs distribution may be problematic or difficult to rigorously define. For a KMS state with temperature $T_\text{KMS}$, the corresponding Wightman function will be periodic in the imaginary time, 
\begin{align}
    \mathcal{W}(\tau - i /T_\text{KMS}, \tau' ) &= \mathcal{W}(\tau', \tau)
\end{align}
where we have utilized the shorthand $\mathcal{W}(\mathsf{x}(\tau), \mathsf{x}'(\tau') ) = \mathcal{W}(\tau , \tau')$. Operationally, a detector that thermalizes with the field will satisfy the KMS detailed balance criterion, commonly stated in the form 
\begin{align}
    \mathcal{R}(\Omega) := \frac{\mathcal{P}_D(\Omega)}{\mathcal{P}_D(-\Omega)} &= e^{-2\pi\Omega/\kappa },
\end{align}
where we have defined $\mathcal{R}(\Omega)$ as the excitation-to-deexcitation ratio of the detector.

\section{Transition probabilities}\label{sec:IV}
\subsection{Gaussian switching}
Using the local and nonlocal Wightman functions, we can calculate the transition probability of the detector in the various superposition configurations of interest. The first approach we consider is a Gaussian switching function centred at the temporal origin of each trajectory,
\begin{align}
    \eta_i(\tau) &= \exp ( - \tau^2/2\sigma_i^2 )
\end{align}
where $\sigma_i$ is a characteristic width for the interaction. Semi-analytic results can be obtained for $\mathcal{P}_D$, following a similar approach to \cite{salton2015acceleration}. We assume a narrowband interaction ($\sigma \ll \kappa ^{-1})$, which allows us to invoke the saddle-point approximation to simplify the double integrals in Eq.\ (\ref{probability}) \cite{nambu2013entanglement,salton2015acceleration}. This regime corresponds with low temperatures (e.g.\ a slow expansion rate in de Sitter). We obtain the following expressions for the transition probability of the detector \cite{foo2020unruhdewitt},
\begin{align}
    \mathcal{P}_T &= \frac{\mathcal{P}_D}{2}  + \frac{2\mathcal{F}_0}{\kappa L}\text{Re}\left( \coth \left( \frac{\kappa }{2}(L+2i\sigma^2\Omega) \right) \right) \vphantom{\bigg]} \\
    \mathcal{P}_\text{dS} &= \mathcal{P}_P = \frac{\mathcal{P}_D}{2} + \frac{\mathcal{F}_0}{(\kappa L/2)^2+\sin^2(\beta)} \vphantom{\bigg]} 
\end{align}
where we have defined
\begin{align}
    \mathcal{P}_D &= \frac{2\mathcal{F}_0}{\sin^2(\beta)}, \vphantom{\frac{e^{-\sigma^2\Omega^2}}{16\pi}} \\ \mathcal{F}_0 &= \frac{(\kappa\sigma\lambda)^2e^{-\sigma^2\Omega^2}}{16\pi}, \vphantom{\frac{e^{-\sigma^2\Omega^2}}{16\pi}} \\
    \beta &= \kappa \sigma^2 \Omega . \vphantom{\frac{e^{-\sigma^2\Omega^2}}{16\pi}}
\end{align}
Note that $\mathcal{P}_D$ is the transition probability of a single detector detecting thermal radiation at the temperature $\kappa (2\pi)^{-1}$ \cite{fooPhysRevD.102.085013,salton2015acceleration}. 
\begin{figure}[h]
    \centering
    \includegraphics[width=0.7\linewidth]{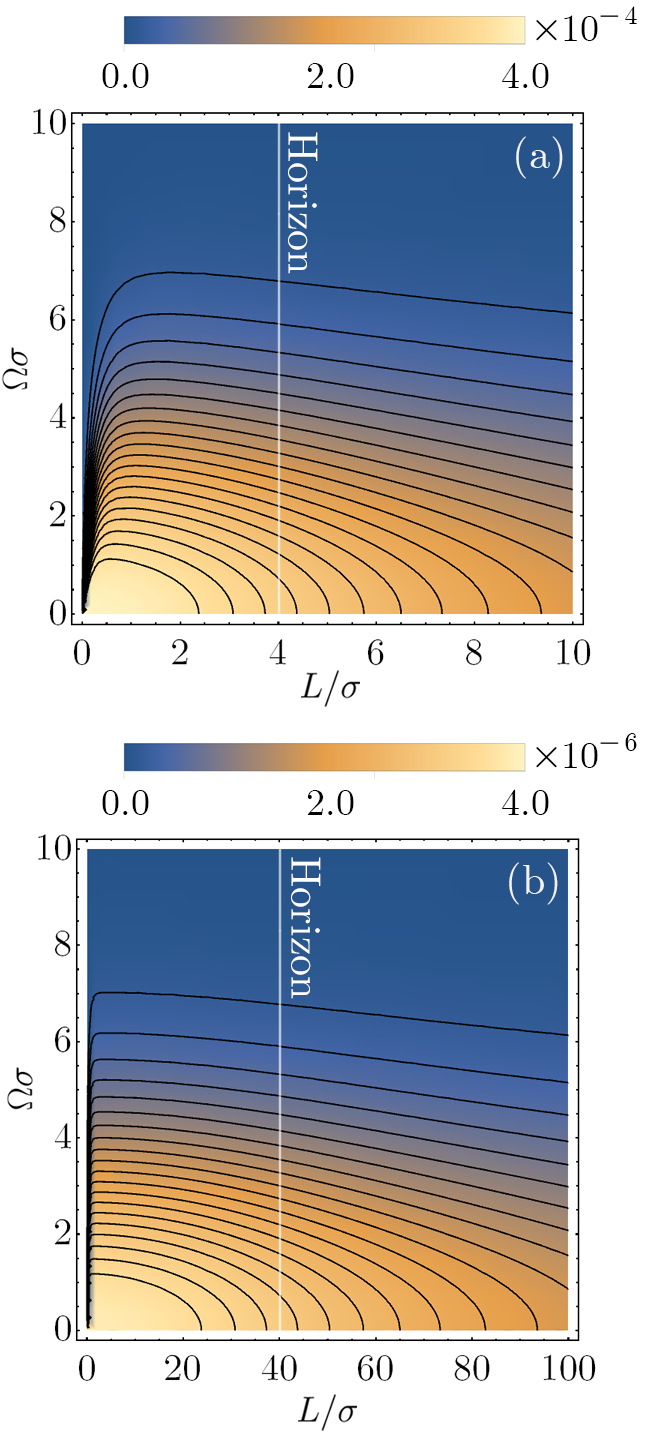}
    \caption{Contour plots of $(\mathcal{P}_T - \mathcal{P}_\text{dS})/\lambda^2$ for (a) $(\sigma\kappa)^{-1} = 4$, and (b) $(\sigma\kappa)^{-1} = 40$. From both plots, we find that the difference in the transition probability between the two spacetimes is largest at small energy gaps. Notably, the interaction regions in the de Sitter spacetime need not be spacelike (beyond the cosmological horizon) in order for the two cases to be distinguished via the respective transition probabilities.}
    \label{fig:diff}
\end{figure}

Several noteworthy observations can be made. First, the nonlocal terms in the transition probability for each scenario vanish in the limit of infinite separation, $L \to \infty$, between the superposed trajectories. This is a generic property of long-range quantum correlations, which decay with distance. In this limit, $\mathcal{P}_D$ reduces to half of that for a single detector in all cases. Second, the presence of the interference term enables the detector to distinguish the thermal bath from the expanding de Sitter spacetime (Fig.\ \ref{fig:diff}), a distinction otherwise inaccessible for a detector traversing a single, classical trajectory. Furthermore, this difference is  {discernible} for small energy gaps and when $L$ is smaller than the cosmological horizon, $L_\text{dS} = (\sigma\kappa)^{-1}$. This contrasts with results found in entanglement harvesting protocols utilizing two detectors travelling on either of the individual trajectories and interacting locally with the field \cite{ver2009entangling,nambu2013entanglement,kukita2017entanglement,tian2014dynamics,tian2016detecting,salton2015acceleration}. Only when the two detectors are separated by a distance larger than $L_\text{dS}$, can the amount of entanglement extracted from the field be used to differentiate these spacetimes. Finally, the transition probability of the detector accelerating along parallel trajectories is identical to the geodesic de Sitter case. While the dynamics are qualitatively different -- an inertial observer measures a constant distance between the accelerated trajectories, whereas the de Sitter trajectories diverge away from each other -- the similarity between them is the constancy of the length scale $L$ (recalling that it is a \textit{co-moving distance} in de Sitter). Note however that this result is obtained under the specific assumption of a narrowband detector-field interaction centred at $\tau = 0$.

\subsection{Compact switching}
Next, we consider an interaction with compact support, with the switching function chosen to be
\begin{align}
    \eta_i(\tau) &= \begin{cases} \cos^2 \left( \frac{\tau - \tau_i}{\sigma_i} \right) & \tau_i - \pi\sigma_i/2 \leq \tau \leq \tau_i + \pi\sigma_i/2 \\
    0 & \text{elsewhere.}
    \end{cases}
\end{align}
This allows us to study the causal relations between localised spacetime regions within which the detector-field interaction occurs
\cite{henderson2020quantum,Cong:2020crf}.

\begin{figure}[h]
    \centering
    \includegraphics[width=0.95\linewidth]{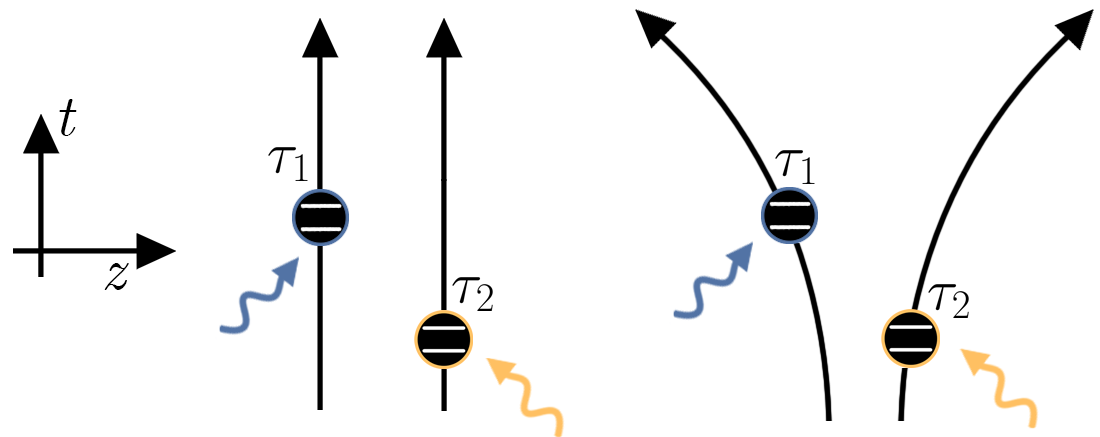}
    \caption{Schematic diagram of the detector trajectories in superposition, with a time-delay between the interaction times on the respective branches. The diagram on the left depicts the thermal bath scenario, while on the right, the diverging geodesics in de Sitter spacetime. In the thermal bath, the interaction regions can always be arranged so that there is some causal contact between them; in the de Sitter case, they will inevitably become spacelike once the interaction regions become separated by the cosmological horizon. }
    \label{fig:my_label}
\end{figure}
\begin{figure*}[t]
    \centering
    \includegraphics[width=1.0\linewidth]{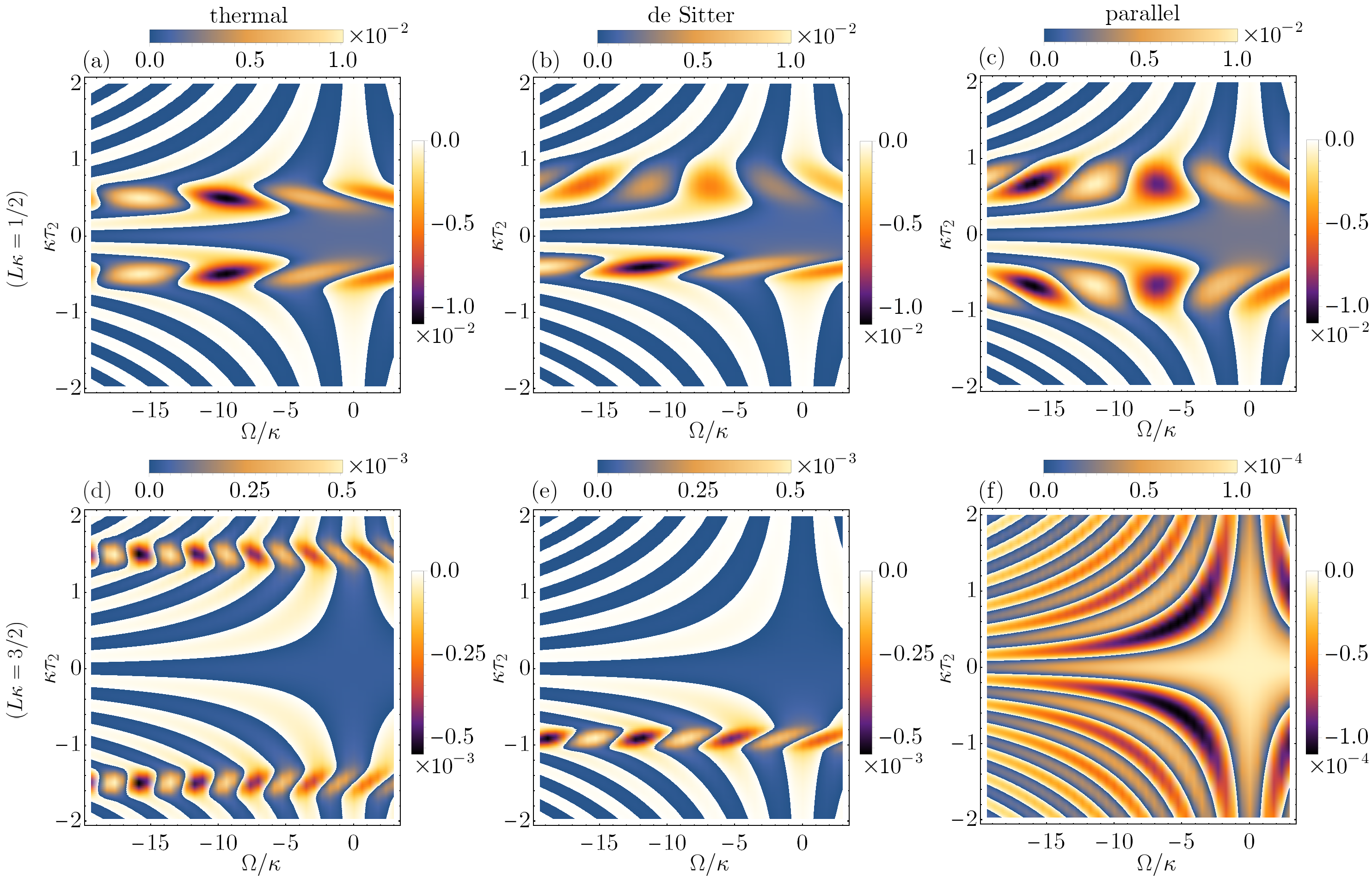}
    \caption{Plots of the interference term, $\sum_{i\neq j}\mathcal{P}_{ij,D}/\lambda^2$, as a function of the energy gap $\Omega/\kappa$ and the centre time of the switching function for the second branch of the superposition, $\kappa\tau_2$. The top row has $L\kappa = 1/2$ while the bottom row has $L \kappa = 3/2$ fixed. The two different color densities used in the plots contrast the regions where $\sum_{i\neq j}\mathcal{P}_{ij,D}/\lambda^2$ is positive and negative; this is highly sensitive to the detector energy gap and the causal relationship between the interaction regions. }
    \label{fig:cosine}
\end{figure*}

Using these switching functions, Eq.\ (\ref{probability}) can be numerically integrated directly. In Fig.\ \ref{fig:cosine}, we have plotted the value of the interference terms, $\sum_{i\neq j} \mathcal{P}_{ij,D}$, in the transition probability and introduced a time-delay between $\tau_1,\tau_2$ (that is, centering one interaction at $\tau_1 = 0$ while varying the central proper time $\tau_2$ along the other trajectory). 
For sufficiently small $L\kappa$, the transition probability displays resonant behaviour at the light-like overlaps of the interaction regions in all three spacetimes. That is, when the interaction region $\eta_{j}(\tau)$ begins to overlap with the light-like extension of the other region $\eta_{i}(\tau)$ (where $\tau_i<\tau_j$), the interference terms either amplify or inhibit transitions in the detector, depending on the energy gap $\Omega$. Above a critical value of $L\kappa$
(and accounting for the interaction width, $\sigma_i$) these resonances are highly suppressed for the parallel accelerated trajectories because the interaction regions are spacelike
for all $\tau_{1},\tau_2$, see Fig.\ \ref{fig:cosine}(f), noting the order of magnitude difference compared with the other cases. In the de Sitter case, only the $\tau_2>0$ resonances disappear since the regions become causally disconnected only after they cross the expansion-induced horizon of the other. This contrasts with the thermal case, where the trajectories always allow for causal contact for some configuration of the interaction regions  {(i.e.\ a larger $L\kappa $ requires a larger time-delay)}. Finally, we note that our use of the terminology `causal' refers to effects that dominate when the {spacetime} regions are in partial or fully lightlike contact.

Since these causal resonances have the ability to suppress the transitions experienced by the detector (i.e.\ the amount of noise perceived by the detector) this suggests that preparing two detectors, each in a superposition of paths and switching times, may enhance their ability to become entangled. This is because bipartite detector entanglement is commonly quantified by the concurrence, $\mathcal{C}_{AB}$, which is effectively a competition between the $\mathcal{M}$ term in Eq.\ (\ref{reduceddensity}), and the geometric mean of the individual transition probabilities:
\begin{align}
    \mathcal{C}_{AB} &:= 2 \text{max} \left[ 0 , | \mathcal{M}|  - \sqrt{\mathcal{P}_A \mathcal{P}_B} \right] .
\end{align}
Indeed, this is corroborated by the results obtained by Henderson et.\ al.\ \cite{henderson2020quantum}, where it was found that the temporal superposition of switching times of two detectors enabled them to become entangled in regimes where it was not possible for detetors with classical switching functions. 

Finally, we note that even when the switching regions on either branch of the superposition are outside the causal cone of the other, there are still interference oscillations which become suppressed as the time-delay becomes very large. This is signatory of the nonlocal correlations between the quantum field degrees of freedom, which exist even between spacelike- and timelike-separated spacetime regions. 

\section{Detector transition rates}\label{sec:V}
Here, we present calculations for the detector transition rate. For the detector superposed at two locations within the thermal bath, both the local and nonlocal Wightman functions satisfy the KMS criterion:
\begin{align}
\begin{split}  
    \mathcal{W}_S(\tau - 2\pi i/\kappa,\tau') &= \mathcal{W}_S(\tau', \tau),  \vphantom{\frac{1}{2}} \\
    \mathcal{W}_T(\tau -2\pi i/\kappa,\tau') &= \mathcal{W}_T(\tau', \tau) .
\end{split} 
\end{align}
Because of this property, we might expect that the detector will exhibit a thermal response to the field. To determine whether the detector does indeed thermalise at the temperature of the quantum field, we evaluate its transition rate. We confirmed numerically that the result is given by 
\begin{align}
    \dot{\mathcal{P}}_D &= \frac{\Omega}{4\pi} \frac{1}{\exp ( 2 \pi\Omega/\kappa) - 1} ( 1 + \text{sinc} ( \Omega L ) ) 
\end{align}
where $\text{sinc}(x) = \sin(x)/x$. Using the identity $\text{sinc}(\Omega L) = \text{sinc}(-\Omega  L)$, we find that the detailed balance form of the KMS criterion is indeed satisfied:
\begin{align}
    \frac{\dot{\mathcal{P}}_D(\Omega)}{\dot{\mathcal{P}}_D(-\Omega)} &= e^{-2\pi \Omega/\kappa} ,
\end{align}
for a thermal field at temperature $\kappa(2\pi)^{-1}$, which confirms the above conjecture. 
\begin{figure}[h]
    \centering
    \includegraphics[width=0.85\linewidth]{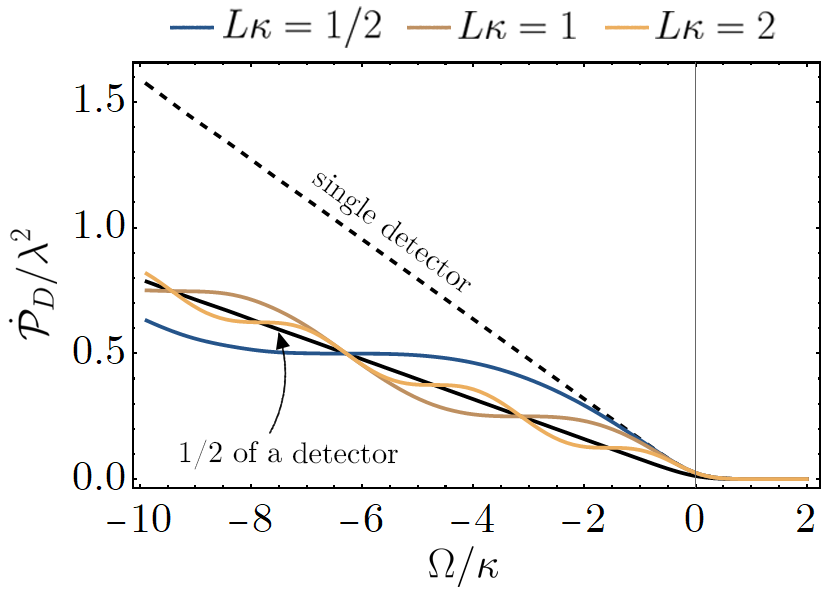}
    \caption{Transition rate of the detector superposed on two paths separated by the constant distance $L$ in a thermal bath in Minkowski spacetime, as a function of the energy gap. The transition rate experiences resonant oscillations as the energy gap is varied; when all of the solid lines intersect, the response is half of that of a detector on a single path. }
    \label{fig:thermaltransitionrate}
\end{figure}
In Fig.\ \ref{fig:thermaltransitionrate}, we have plotted the transition rate of the detector as a function of the energy gap. Upon introducing the superposition of paths, the transition rate oscillates with increasingly negative energy gaps, with the frequency of these oscillations increasing with the path separation. The interference term vanishes at periodic values of the energy gap, indicating that the response of the detector is both thermal and Planckian for those values. 

The time-translation invariant thermal response of the detector in the heat bath contrasts both the co-moving de Sitter superposition and the parallel acceleration superposition, where the nonlocal Wightman functions are dependent on the sum of the proper times between the two trajectories, i.e.\ they are time-dependent. Note that the latter case is studied in \cite{fooPhysRevD.102.085013}, so we focus here on the quantum-controlled detector transition rate in de Sitter spacetime.

The time-dependence of the nonlocal Wightman functions indicates that the correlation structure of the quantum field depends strongly on the different spacetime regions that the detector probes along its trajectory. Moreover, the transition rate will generally not satisfy the detailed balance criterion:
\begin{align}
    \frac{\dot{\mathcal{P}}_D(\Omega)}{\dot{\mathsf{P}}_D(-\Omega)} &= \frac{\dot{\mathcal{P}}_{ii,D}(\Omega) + \dot{\mathcal{P}}_{ij,D}(\Omega)}{\dot{\mathcal{P}}_{ii,D}(-\Omega) + \dot{\mathcal{P}}_{ij,D}(-\Omega)} \neq e^{-2\pi \Omega/\kappa} 
\end{align}
since the interference terms are generally asymmetric in $\Omega$. This is intriguing because the paths would individually elicit a thermal response in the detector at an identical temperature, yet superposing the detector along those paths yields a nonthermal response. 
\begin{figure}[h]
    \centering
    \includegraphics[width=0.825\linewidth]{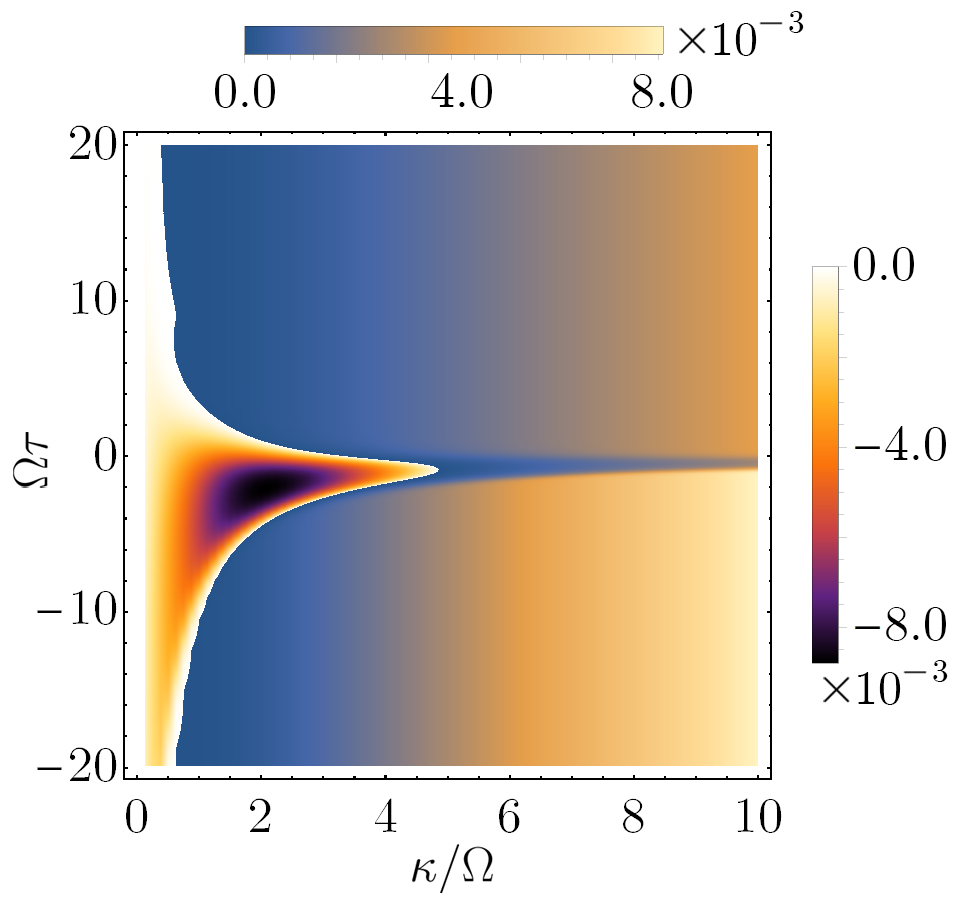}
    \caption{Transition rate, $\dot{\mathcal{P}}_D/\lambda^2$, of the detector in a superposition of two geodesic paths in de Sitter spacetime, as a function of the dimensionless proper time and expansion rate. The regions shaded blue to light yellow correspond to positive transition rates, while the regions shaded dark purple to white correspond to negative transition rate. For sufficiently high expansion rates, these negative regions vanish. }
    \label{fig:desitter1}
\end{figure}
In Fig.\ \ref{fig:desitter1} we have plotted the transition rate of a detector superposed along two paths in the de Sitter universe separated by a constant co-moving distance, as a function of the dimensionless expansion rate, $\kappa/\Omega$, and the dimensionless proper time at which the detector (ensemble) is measured, $\Omega \tau$ $(\tau_F \equiv \tau)$. In the asymptotic past, the detector exhibits the same thermal response that a single detector would experience for all times. This can be seen from the past asymptotic form of the nonlocal Wightman function, which approaches that of a single detector as $\tau \to -\infty$:
\begin{align}
    \lim_{\tau\to - \infty} \mathcal{W}_\text{dS}(p,s) = \mathcal{W}_D(s).
\end{align}
Near $\tau = 0$, the trajectories begin to bifurcate from each other, and the transition rate dips before equilibrating towards that of half of a single detector into as $\tau \to \infty$.
\begin{figure}[h]
    \centering
    \includegraphics[width=0.925\linewidth]{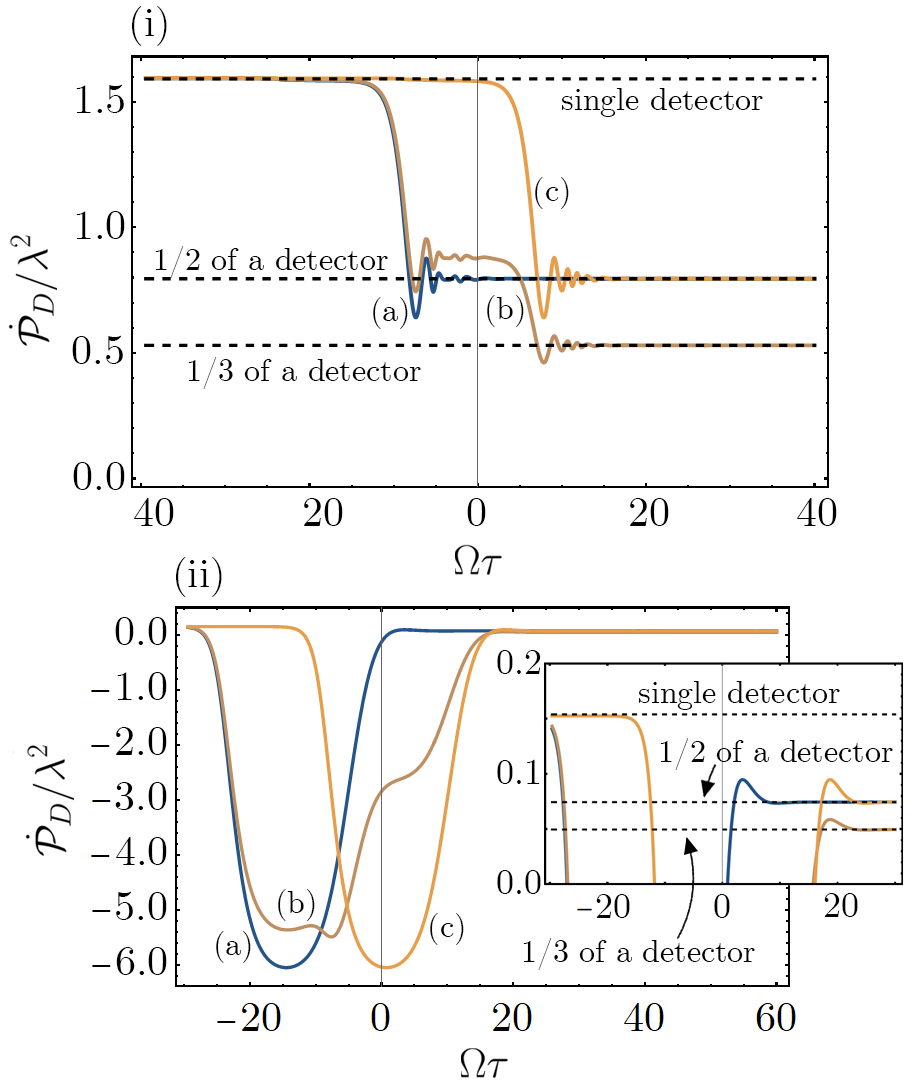}
    \caption{(i) Transition rate of a detector with negative energy gap, $\Omega/\kappa = -10$, in a superposition of three geodesic paths, with the respective separations $L_{12} = 0.01$ and $L_{13} = 20$. The respective lines in the plot show (a) transition rate of the detector in a superposition of co-moving distance $L_{13}$, (b) transition rate of the detector in a superposition of three co-moving paths with separation $L_{12}$, $L_{13}$ and $L_{23}$ and (c) the transition rate of the detector in a superposition of two paths separated by the co-moving distance $L_{12}$. In the regions where the structure of the field correlations changes rapidly due to the diverging Euclidean distance between the superposed paths, the total transition oscillates, eventually equilibrating to a third of that experienced by a single detector. (ii) The same setup for a detector with a positive energy gap, $\Omega/\kappa = 0.5$. The inset shows a zoomed in view of the transition rate. }
    \label{fig:desitter2}
\end{figure}
Interestingly, for sufficiently slow expansion rates, near the proper time origin the detector's transition rate becomes negative before becoming positive again as the superposed paths recede away from each other. This region of negative transition rate is consistent with the ensemble interpretation mentioned previously; other works have also noted the presence of such regions in a variety of spacetimes \cite{Louko:2007mu,Smith:2013zqa,langlois2006causal}. For our quantum-controlled detector, the regimes of negative transition rate correspond to spacetime regions in which the correlation structure between the two paths is strongly time-dependent, inducing stronger interference between the paths. Finally, since the detector is switched on in the asymptotic past and has an effectively constant transition rate until the region of strong interference, an integral of the transition rate with respect to the proper time will always yields a positive result, as required by a physical meaning of this quantity as a transition probability. (This remains true if one considers smooth switching functions.)  

In Fig.\ \ref{fig:desitter2}, we have plotted the transition rate for a detector in a superposition of three geodesic paths, for (top) negative and (bottom) positive energy gaps. We have chosen a setup such that two of the paths have a small co-moving distance between them $(L_{12} \ll \kappa^{-1})$, which are separated by a large distance from the third trajectory $(L_{13}$, $L_{23} \gg \kappa^{-1}$). For negative gaps, the detector response is approximately time-independent and thermal, until the point at which the third trajectory begins to recede from the first two. The second dip occurs when the first pair of trajectories recede from each other, after which the transition rate approaches that of one-third of a single detector. For positive gaps, we observe similar regions of behaviour, where the aforementioned period of negative transition rate is manifest.

\section{Conclusion}\label{sec:VI}
We have shown that by introducing a quantum-controlled superposition of trajectories, a UdW detector gains information about the field and the global structure of spacetime through nonlocal correlation functions that would be otherwise inaccessible to a single detector. In particular, such a detector can discriminate between the thermal state of a field in Minkowski spacetime and a vacuum state of an exponentially expanding de Sitter universe, which is proven to be impossible for a single detector traversing a classical trajectory \cite{gibbons1977cosmological}, and only achievable by examining the amount entanglement between two classically moving detectors when their spatial separation is larger than the cosmological event horizon \cite{ver2009entangling,nambu2013entanglement,kukita2017entanglement,tian2014dynamics,tian2016detecting,salton2015acceleration}. The interference effects illustrate the rich nonlocal features of quantum fields, and how they depend strongly on the dynamical properties of the trajectories traversed by the detector in superposition. Because of this, one would expect such detectors can be utilised in the study of foundational questions about causality in quantum theory and the geometric structure of spacetime from the perspective of relativistic and curved spacetime QFT. 

Our quantum-controlled detector model represents a concrete physical realisation of a `superposition' of quantum channels (unitaries) acting on a quantum system \cite{oi2003interference,chiribella2019quantum}, so far studied in  abstract quantum information and foundations settings, for instance, quantum communication~\cite{guerinPhysRevLett.117.100502,chiribellaPhysRevA.86.040301,Abbott_2020}, causality~\cite{oreshkov2012quantum,eblerPhysRevLett.120.120502} and thermodynamics~\cite{felcePhysRevLett.125.070603}. Furthermore, our detector model can be applied to study in a novel way the nature of time and decoherence in gravitational interferometry setups, systems which have received renewed interest as testbeds of quantum-gravitational physics \cite{zych2019bell,Zych:2011hu,Zych:2012ut,Smith:2019imm}. Our study thus builds a new, direct connection between these fields of research and relativistic quantum field theory in curved spacetime. Finally, our here studied quantum extension of a traditional detector model also allows us to build a bottom-up approach for studying the operational effects produced by quantum superpositions of classical spacetime geometries \cite{ford1997cosmological,lake2019generalised,bose2017spin,christodoulou2019possibility,marletto2017gravitationally}. 

\section{Acknowledgements}
M.Z. acknowledges support from ARC grants EQuS CE170100009 and DECRA DE180101443. R.B.M acknowledges support from the Natural Sciences and Engineering Research Council of Canada and from AOARD Grant FA2386-19-1-4077.

\bibliography{References.bib}

\begin{thebibliography}{60}%
\makeatletter
\providecommand \@ifxundefined [1]{%
 \@ifx{#1\undefined}
}%
\providecommand \@ifnum [1]{%
 \ifnum #1\expandafter \@firstoftwo
 \else \expandafter \@secondoftwo
 \fi
}%
\providecommand \@ifx [1]{%
 \ifx #1\expandafter \@firstoftwo
 \else \expandafter \@secondoftwo
 \fi
}%
\providecommand \natexlab [1]{#1}%
\providecommand \enquote  [1]{``#1''}%
\providecommand \bibnamefont  [1]{#1}%
\providecommand \bibfnamefont [1]{#1}%
\providecommand \citenamefont [1]{#1}%
\providecommand \href@noop [0]{\@secondoftwo}%
\providecommand \href [0]{\begingroup \@sanitize@url \@href}%
\providecommand \@href[1]{\@@startlink{#1}\@@href}%
\providecommand \@@href[1]{\endgroup#1\@@endlink}%
\providecommand \@sanitize@url [0]{\catcode `\\12\catcode `\$12\catcode
  `\&12\catcode `\#12\catcode `\^12\catcode `\_12\catcode `\%12\relax}%
\providecommand \@@startlink[1]{}%
\providecommand \@@endlink[0]{}%
\providecommand \url  [0]{\begingroup\@sanitize@url \@url }%
\providecommand \@url [1]{\endgroup\@href {#1}{\urlprefix }}%
\providecommand \urlprefix  [0]{URL }%
\providecommand \Eprint [0]{\href }%
\providecommand \doibase [0]{https://doi.org/}%
\providecommand \selectlanguage [0]{\@gobble}%
\providecommand \bibinfo  [0]{\@secondoftwo}%
\providecommand \bibfield  [0]{\@secondoftwo}%
\providecommand \translation [1]{[#1]}%
\providecommand \BibitemOpen [0]{}%
\providecommand \bibitemStop [0]{}%
\providecommand \bibitemNoStop [0]{.\EOS\space}%
\providecommand \EOS [0]{\spacefactor3000\relax}%
\providecommand \BibitemShut  [1]{\csname bibitem#1\endcsname}%
\let\auto@bib@innerbib\@empty
\bibitem [{\citenamefont {Sorkin}(1993)}]{Sorkin:1993gg}%
  \BibitemOpen
  \bibfield  {author} {\bibinfo {author} {\bibfnamefont {R.~D.}\ \bibnamefont
  {Sorkin}},\ }\bibfield  {title} {\bibinfo {title} {{Impossible measurements
  on quantum fields}}\ }(\bibinfo {year} {1993})\ \Eprint
  {https://arxiv.org/abs/gr-qc/9302018} {arXiv:gr-qc/9302018} \BibitemShut
  {NoStop}%
\bibitem [{\citenamefont {Martin}\ and\ \citenamefont
  {Schwinger}(1959)}]{martinPhysRev.115.1342}%
  \BibitemOpen
  \bibfield  {author} {\bibinfo {author} {\bibfnamefont {P.~C.}\ \bibnamefont
  {Martin}}\ and\ \bibinfo {author} {\bibfnamefont {J.}~\bibnamefont
  {Schwinger}},\ }\bibfield  {title} {\bibinfo {title} {Theory of many-particle
  systems. i},\ }\href {https://doi.org/10.1103/PhysRev.115.1342} {\bibfield
  {journal} {\bibinfo  {journal} {Phys. Rev.}\ }\textbf {\bibinfo {volume}
  {115}},\ \bibinfo {pages} {1342} (\bibinfo {year} {1959})}\BibitemShut
  {NoStop}%
\bibitem [{\citenamefont {de~Ram\'on}\ \emph {et~al.}(2021)\citenamefont
  {de~Ram\'on}, \citenamefont {Papageorgiou},\ and\ \citenamefont
  {Mart\'{\i}n-Mart\'{\i}nez}}]{ramonPhysRevD.103.085002}%
  \BibitemOpen
  \bibfield  {author} {\bibinfo {author} {\bibfnamefont {J.}~\bibnamefont
  {de~Ram\'on}}, \bibinfo {author} {\bibfnamefont {M.}~\bibnamefont
  {Papageorgiou}},\ and\ \bibinfo {author} {\bibfnamefont {E.}~\bibnamefont
  {Mart\'{\i}n-Mart\'{\i}nez}},\ }\bibfield  {title} {\bibinfo {title}
  {Relativistic causality in particle detector models: Faster-than-light
  signaling and impossible measurements},\ }\href
  {https://doi.org/10.1103/PhysRevD.103.085002} {\bibfield  {journal} {\bibinfo
   {journal} {Phys. Rev. D}\ }\textbf {\bibinfo {volume} {103}},\ \bibinfo
  {pages} {085002} (\bibinfo {year} {2021})}\BibitemShut {NoStop}%
\bibitem [{\citenamefont {Funai}\ and\ \citenamefont
  {Mart\'{\i}n-Mart\'{\i}nez}(2019)}]{funaiPhysRevD.100.065021}%
  \BibitemOpen
  \bibfield  {author} {\bibinfo {author} {\bibfnamefont {N.}~\bibnamefont
  {Funai}}\ and\ \bibinfo {author} {\bibfnamefont {E.}~\bibnamefont
  {Mart\'{\i}n-Mart\'{\i}nez}},\ }\bibfield  {title} {\bibinfo {title}
  {Faster-than-light signaling in the rotating-wave approximation},\ }\href
  {https://doi.org/10.1103/PhysRevD.100.065021} {\bibfield  {journal} {\bibinfo
   {journal} {Phys. Rev. D}\ }\textbf {\bibinfo {volume} {100}},\ \bibinfo
  {pages} {065021} (\bibinfo {year} {2019})}\BibitemShut {NoStop}%
\bibitem [{\citenamefont {Fewster}\ and\ \citenamefont
  {Verch}(2020)}]{Fewster:2018qbm}%
  \BibitemOpen
  \bibfield  {author} {\bibinfo {author} {\bibfnamefont {C.~J.}\ \bibnamefont
  {Fewster}}\ and\ \bibinfo {author} {\bibfnamefont {R.}~\bibnamefont
  {Verch}},\ }\bibfield  {title} {\bibinfo {title} {{Quantum fields and local
  measurements}},\ }\href {https://doi.org/10.1007/s00220-020-03800-6}
  {\bibfield  {journal} {\bibinfo  {journal} {Commun. Math. Phys.}\ }\textbf
  {\bibinfo {volume} {378}},\ \bibinfo {pages} {851} (\bibinfo {year}
  {2020})},\ \Eprint {https://arxiv.org/abs/1810.06512} {arXiv:1810.06512
  [math-ph]} \BibitemShut {NoStop}%
\bibitem [{\citenamefont {Unruh}\ and\ \citenamefont
  {Wald}(1984)}]{unruh1984happens}%
  \BibitemOpen
  \bibfield  {author} {\bibinfo {author} {\bibfnamefont {W.~G.}\ \bibnamefont
  {Unruh}}\ and\ \bibinfo {author} {\bibfnamefont {R.~M.}\ \bibnamefont
  {Wald}},\ }\bibfield  {title} {\bibinfo {title} {What happens when an
  accelerating observer detects a rindler particle},\ }\href
  {https://doi.org/10.1103/PhysRevD.29.1047} {\bibfield  {journal} {\bibinfo
  {journal} {Phys. Rev. D}\ }\textbf {\bibinfo {volume} {29}},\ \bibinfo
  {pages} {1047} (\bibinfo {year} {1984})}\BibitemShut {NoStop}%
\bibitem [{\citenamefont {Davies}\ and\ \citenamefont
  {Ottewill}(2002)}]{davies2002detection}%
  \BibitemOpen
  \bibfield  {author} {\bibinfo {author} {\bibfnamefont {P.~C.~W.}\
  \bibnamefont {Davies}}\ and\ \bibinfo {author} {\bibfnamefont {A.~C.}\
  \bibnamefont {Ottewill}},\ }\bibfield  {title} {\bibinfo {title} {Detection
  of negative energy: 4-dimensional examples},\ }\href
  {https://doi.org/10.1103/PhysRevD.65.104014} {\bibfield  {journal} {\bibinfo
  {journal} {Phys. Rev. D}\ }\textbf {\bibinfo {volume} {65}},\ \bibinfo
  {pages} {104014} (\bibinfo {year} {2002})}\BibitemShut {NoStop}%
\bibitem [{\citenamefont {Birrell}\ and\ \citenamefont
  {Davies}(1984)}]{birrell1984quantum}%
  \BibitemOpen
  \bibfield  {author} {\bibinfo {author} {\bibfnamefont {N.~D.}\ \bibnamefont
  {Birrell}}\ and\ \bibinfo {author} {\bibfnamefont {P.}~\bibnamefont
  {Davies}},\ }\href@noop {} {\emph {\bibinfo {title} {Quantum fields in curved
  space}}},\ \bibinfo {number} {7}\ (\bibinfo  {publisher} {Cambridge
  university press},\ \bibinfo {year} {1984})\BibitemShut {NoStop}%
\bibitem [{\citenamefont {Satz}(2007)}]{satz2007then}%
  \BibitemOpen
  \bibfield  {author} {\bibinfo {author} {\bibfnamefont {A.}~\bibnamefont
  {Satz}},\ }\bibfield  {title} {\bibinfo {title} {Then again, how often does
  the unruh{\textendash}{DeWitt} detector click if we switch it carefully?},\
  }\href {https://doi.org/10.1088/0264-9381/24/7/003} {\bibfield  {journal}
  {\bibinfo  {journal} {Classical and Quantum Gravity}\ }\textbf {\bibinfo
  {volume} {24}},\ \bibinfo {pages} {1719} (\bibinfo {year}
  {2007})}\BibitemShut {NoStop}%
\bibitem [{\citenamefont {Louko}\ and\ \citenamefont
  {Satz}(2006)}]{louko2006often}%
  \BibitemOpen
  \bibfield  {author} {\bibinfo {author} {\bibfnamefont {J.}~\bibnamefont
  {Louko}}\ and\ \bibinfo {author} {\bibfnamefont {A.}~\bibnamefont {Satz}},\
  }\bibfield  {title} {\bibinfo {title} {How often does the
  unruh{\textendash}{DeWitt} detector click? regularization by a spatial
  profile},\ }\href {https://doi.org/10.1088/0264-9381/23/22/015} {\bibfield
  {journal} {\bibinfo  {journal} {Classical and Quantum Gravity}\ }\textbf
  {\bibinfo {volume} {23}},\ \bibinfo {pages} {6321} (\bibinfo {year}
  {2006})}\BibitemShut {NoStop}%
\bibitem [{\citenamefont {Lin}\ and\ \citenamefont
  {Hu}(2007)}]{lin2007backreaction}%
  \BibitemOpen
  \bibfield  {author} {\bibinfo {author} {\bibfnamefont {S.-Y.}\ \bibnamefont
  {Lin}}\ and\ \bibinfo {author} {\bibfnamefont {B.~L.}\ \bibnamefont {Hu}},\
  }\bibfield  {title} {\bibinfo {title} {Backreaction and the unruh effect: New
  insights from exact solutions of uniformly accelerated detectors},\ }\href
  {https://doi.org/10.1103/PhysRevD.76.064008} {\bibfield  {journal} {\bibinfo
  {journal} {Phys. Rev. D}\ }\textbf {\bibinfo {volume} {76}},\ \bibinfo
  {pages} {064008} (\bibinfo {year} {2007})}\BibitemShut {NoStop}%
\bibitem [{\citenamefont {Sriramkumar}\ and\ \citenamefont
  {Padmanabhan}(1996)}]{sriramkumar1996finite}%
  \BibitemOpen
  \bibfield  {author} {\bibinfo {author} {\bibfnamefont {L.}~\bibnamefont
  {Sriramkumar}}\ and\ \bibinfo {author} {\bibfnamefont {T.}~\bibnamefont
  {Padmanabhan}},\ }\bibfield  {title} {\bibinfo {title} {Finite-time response
  of inertial and uniformly accelerated unruh - {DeWitt} detectors},\ }\href
  {https://doi.org/10.1088/0264-9381/13/8/005} {\bibfield  {journal} {\bibinfo
  {journal} {Classical and Quantum Gravity}\ }\textbf {\bibinfo {volume}
  {13}},\ \bibinfo {pages} {2061} (\bibinfo {year} {1996})}\BibitemShut
  {NoStop}%
\bibitem [{\citenamefont {Louko}\ and\ \citenamefont
  {Satz}(2008{\natexlab{a}})}]{louko2008transition}%
  \BibitemOpen
  \bibfield  {author} {\bibinfo {author} {\bibfnamefont {J.}~\bibnamefont
  {Louko}}\ and\ \bibinfo {author} {\bibfnamefont {A.}~\bibnamefont {Satz}},\
  }\bibfield  {title} {\bibinfo {title} {Transition rate of the unruh–dewitt
  detector in curved spacetime},\ }\href
  {https://doi.org/10.1088/0264-9381/25/5/055012} {\bibfield  {journal}
  {\bibinfo  {journal} {Classical and Quantum Gravity}\ }\textbf {\bibinfo
  {volume} {25}},\ \bibinfo {pages} {055012} (\bibinfo {year}
  {2008}{\natexlab{a}})}\BibitemShut {NoStop}%
\bibitem [{\citenamefont {Hodgkinson}\ and\ \citenamefont
  {Louko}(2012)}]{hodgkinson2012static}%
  \BibitemOpen
  \bibfield  {author} {\bibinfo {author} {\bibfnamefont {L.}~\bibnamefont
  {Hodgkinson}}\ and\ \bibinfo {author} {\bibfnamefont {J.}~\bibnamefont
  {Louko}},\ }\bibfield  {title} {\bibinfo {title} {{Static, stationary and
  inertial Unruh-DeWitt detectors on the BTZ black hole}},\ }\href
  {https://doi.org/10.1103/PhysRevD.86.064031} {\bibfield  {journal} {\bibinfo
  {journal} {Phys. Rev. D}\ }\textbf {\bibinfo {volume} {86}},\ \bibinfo
  {pages} {064031} (\bibinfo {year} {2012})},\ \Eprint
  {https://arxiv.org/abs/1206.2055} {arXiv:1206.2055 [gr-qc]} \BibitemShut
  {NoStop}%
\bibitem [{\citenamefont {Ng}\ \emph {et~al.}(2014)\citenamefont {Ng},
  \citenamefont {Hodgkinson}, \citenamefont {Louko}, \citenamefont {Mann},\
  and\ \citenamefont {Martin-Martinez}}]{ng2014unruh}%
  \BibitemOpen
  \bibfield  {author} {\bibinfo {author} {\bibfnamefont {K.~K.}\ \bibnamefont
  {Ng}}, \bibinfo {author} {\bibfnamefont {L.}~\bibnamefont {Hodgkinson}},
  \bibinfo {author} {\bibfnamefont {J.}~\bibnamefont {Louko}}, \bibinfo
  {author} {\bibfnamefont {R.~B.}\ \bibnamefont {Mann}},\ and\ \bibinfo
  {author} {\bibfnamefont {E.}~\bibnamefont {Martin-Martinez}},\ }\bibfield
  {title} {\bibinfo {title} {{Unruh-DeWitt detector response along static and
  circular geodesic trajectories for Schwarzschild-AdS black holes}},\ }\href
  {https://doi.org/10.1103/PhysRevD.90.064003} {\bibfield  {journal} {\bibinfo
  {journal} {Phys. Rev. D}\ }\textbf {\bibinfo {volume} {90}},\ \bibinfo
  {pages} {064003} (\bibinfo {year} {2014})},\ \Eprint
  {https://arxiv.org/abs/1406.2688} {arXiv:1406.2688 [quant-ph]} \BibitemShut
  {NoStop}%
\bibitem [{\citenamefont {Ng}\ \emph {et~al.}(2017)\citenamefont {Ng},
  \citenamefont {Mann},\ and\ \citenamefont
  {Mart\'{\i}n-Mart\'{\i}nez}}]{ng2017over}%
  \BibitemOpen
  \bibfield  {author} {\bibinfo {author} {\bibfnamefont {K.~K.}\ \bibnamefont
  {Ng}}, \bibinfo {author} {\bibfnamefont {R.~B.}\ \bibnamefont {Mann}},\ and\
  \bibinfo {author} {\bibfnamefont {E.}~\bibnamefont
  {Mart\'{\i}n-Mart\'{\i}nez}},\ }\bibfield  {title} {\bibinfo {title} {Over
  the horizon: Distinguishing the schwarzschild spacetime and the
  $\mathbb{R}{\mathbb{p}}^{3}$ spacetime using an unruh-dewitt detector},\
  }\href {https://doi.org/10.1103/PhysRevD.96.085004} {\bibfield  {journal}
  {\bibinfo  {journal} {Phys. Rev. D}\ }\textbf {\bibinfo {volume} {96}},\
  \bibinfo {pages} {085004} (\bibinfo {year} {2017})}\BibitemShut {NoStop}%
\bibitem [{\citenamefont {Unruh}(1976)}]{unruh1976notes}%
  \BibitemOpen
  \bibfield  {author} {\bibinfo {author} {\bibfnamefont {W.~G.}\ \bibnamefont
  {Unruh}},\ }\bibfield  {title} {\bibinfo {title} {Notes on black-hole
  evaporation},\ }\href {https://doi.org/10.1103/PhysRevD.14.870} {\bibfield
  {journal} {\bibinfo  {journal} {Phys. Rev. D}\ }\textbf {\bibinfo {volume}
  {14}},\ \bibinfo {pages} {870} (\bibinfo {year} {1976})}\BibitemShut
  {NoStop}%
\bibitem [{\citenamefont {Gibbons}\ and\ \citenamefont
  {Hawking}(1977)}]{gibbons1977cosmological}%
  \BibitemOpen
  \bibfield  {author} {\bibinfo {author} {\bibfnamefont {G.~W.}\ \bibnamefont
  {Gibbons}}\ and\ \bibinfo {author} {\bibfnamefont {S.~W.}\ \bibnamefont
  {Hawking}},\ }\bibfield  {title} {\bibinfo {title} {Cosmological event
  horizons, thermodynamics, and particle creation},\ }\href
  {https://doi.org/10.1103/PhysRevD.15.2738} {\bibfield  {journal} {\bibinfo
  {journal} {Phys. Rev. D}\ }\textbf {\bibinfo {volume} {15}},\ \bibinfo
  {pages} {2738} (\bibinfo {year} {1977})}\BibitemShut {NoStop}%
\bibitem [{\citenamefont {Foo}\ \emph {et~al.}(2020{\natexlab{a}})\citenamefont
  {Foo}, \citenamefont {Onoe},\ and\ \citenamefont
  {Zych}}]{fooPhysRevD.102.085013}%
  \BibitemOpen
  \bibfield  {author} {\bibinfo {author} {\bibfnamefont {J.}~\bibnamefont
  {Foo}}, \bibinfo {author} {\bibfnamefont {S.}~\bibnamefont {Onoe}},\ and\
  \bibinfo {author} {\bibfnamefont {M.}~\bibnamefont {Zych}},\ }\bibfield
  {title} {\bibinfo {title} {Unruh-dewitt detectors in quantum superpositions
  of trajectories},\ }\href {https://doi.org/10.1103/PhysRevD.102.085013}
  {\bibfield  {journal} {\bibinfo  {journal} {Phys. Rev. D}\ }\textbf {\bibinfo
  {volume} {102}},\ \bibinfo {pages} {085013} (\bibinfo {year}
  {2020}{\natexlab{a}})}\BibitemShut {NoStop}%
\bibitem [{\citenamefont {Barbado}\ \emph {et~al.}(2020)\citenamefont
  {Barbado}, \citenamefont {Castro-Ruiz}, \citenamefont {Apadula},\ and\
  \citenamefont {Brukner}}]{barbadoPhysRevD.102.045002}%
  \BibitemOpen
  \bibfield  {author} {\bibinfo {author} {\bibfnamefont {L.~C.}\ \bibnamefont
  {Barbado}}, \bibinfo {author} {\bibfnamefont {E.}~\bibnamefont
  {Castro-Ruiz}}, \bibinfo {author} {\bibfnamefont {L.}~\bibnamefont
  {Apadula}},\ and\ \bibinfo {author} {\bibfnamefont {i.~c.~v.}\ \bibnamefont
  {Brukner}},\ }\bibfield  {title} {\bibinfo {title} {Unruh effect for
  detectors in superposition of accelerations},\ }\href
  {https://doi.org/10.1103/PhysRevD.102.045002} {\bibfield  {journal} {\bibinfo
   {journal} {Phys. Rev. D}\ }\textbf {\bibinfo {volume} {102}},\ \bibinfo
  {pages} {045002} (\bibinfo {year} {2020})}\BibitemShut {NoStop}%
\bibitem [{\citenamefont {Rabochaya}\ and\ \citenamefont
  {Zerbini}(2016)}]{rabochaya2016quantum}%
  \BibitemOpen
  \bibfield  {author} {\bibinfo {author} {\bibfnamefont {Y.}~\bibnamefont
  {Rabochaya}}\ and\ \bibinfo {author} {\bibfnamefont {S.}~\bibnamefont
  {Zerbini}},\ }\bibfield  {title} {\bibinfo {title} {{Quantum detectors in
  generic non flat FLRW space-times}},\ }\href
  {https://doi.org/10.1007/s10773-015-2902-x} {\bibfield  {journal} {\bibinfo
  {journal} {Int. J. Theor. Phys.}\ }\textbf {\bibinfo {volume} {55}},\
  \bibinfo {pages} {2682} (\bibinfo {year} {2016})},\ \Eprint
  {https://arxiv.org/abs/1505.00998} {arXiv:1505.00998 [gr-qc]} \BibitemShut
  {NoStop}%
\bibitem [{\citenamefont {Garbrecht}\ and\ \citenamefont
  {Prokopec}(2004)}]{garbrecht2004unruh}%
  \BibitemOpen
  \bibfield  {author} {\bibinfo {author} {\bibfnamefont {B.}~\bibnamefont
  {Garbrecht}}\ and\ \bibinfo {author} {\bibfnamefont {T.}~\bibnamefont
  {Prokopec}},\ }\bibfield  {title} {\bibinfo {title} {{Unruh response
  functions for scalar fields in de Sitter space}},\ }\href
  {https://doi.org/10.1088/0264-9381/21/21/016} {\bibfield  {journal} {\bibinfo
   {journal} {Class. Quant. Grav.}\ }\textbf {\bibinfo {volume} {21}},\
  \bibinfo {pages} {4993} (\bibinfo {year} {2004})},\ \Eprint
  {https://arxiv.org/abs/gr-qc/0404058} {arXiv:gr-qc/0404058} \BibitemShut
  {NoStop}%
\bibitem [{\citenamefont {Acquaviva}\ \emph {et~al.}(2012)\citenamefont
  {Acquaviva}, \citenamefont {Di~Criscienzo}, \citenamefont {Tolotti},
  \citenamefont {Vanzo},\ and\ \citenamefont {Zerbini}}]{acquaviva2012unruh}%
  \BibitemOpen
  \bibfield  {author} {\bibinfo {author} {\bibfnamefont {G.}~\bibnamefont
  {Acquaviva}}, \bibinfo {author} {\bibfnamefont {R.}~\bibnamefont
  {Di~Criscienzo}}, \bibinfo {author} {\bibfnamefont {M.}~\bibnamefont
  {Tolotti}}, \bibinfo {author} {\bibfnamefont {L.}~\bibnamefont {Vanzo}},\
  and\ \bibinfo {author} {\bibfnamefont {S.}~\bibnamefont {Zerbini}},\
  }\bibfield  {title} {\bibinfo {title} {{Unruh-DeWitt detectors in spherically
  symmetric dynamical space-times}},\ }\href
  {https://doi.org/10.1007/s10773-011-1033-2} {\bibfield  {journal} {\bibinfo
  {journal} {Int. J. Theor. Phys.}\ }\textbf {\bibinfo {volume} {51}},\
  \bibinfo {pages} {1555} (\bibinfo {year} {2012})},\ \Eprint
  {https://arxiv.org/abs/1111.6389} {arXiv:1111.6389 [gr-qc]} \BibitemShut
  {NoStop}%
\bibitem [{\citenamefont {Tian}\ and\ \citenamefont
  {Jing}(2013)}]{tian2013geometric}%
  \BibitemOpen
  \bibfield  {author} {\bibinfo {author} {\bibfnamefont {Z.}~\bibnamefont
  {Tian}}\ and\ \bibinfo {author} {\bibfnamefont {J.}~\bibnamefont {Jing}},\
  }\bibfield  {title} {\bibinfo {title} {Geometric phase of two-level atoms and
  thermal nature of de sitter spacetime},\ }\bibfield  {journal} {\bibinfo
  {journal} {Journal of High Energy Physics}\ }\textbf {\bibinfo {volume}
  {2013}},\ \href {https://doi.org/10.1007/jhep04(2013)109}
  {10.1007/jhep04(2013)109} (\bibinfo {year} {2013})\BibitemShut {NoStop}%
\bibitem [{\citenamefont {Singh}\ \emph {et~al.}(2013)\citenamefont {Singh},
  \citenamefont {Ganguly},\ and\ \citenamefont
  {Padmanabhan}}]{singh2013quantum}%
  \BibitemOpen
  \bibfield  {author} {\bibinfo {author} {\bibfnamefont {S.}~\bibnamefont
  {Singh}}, \bibinfo {author} {\bibfnamefont {C.}~\bibnamefont {Ganguly}},\
  and\ \bibinfo {author} {\bibfnamefont {T.}~\bibnamefont {Padmanabhan}},\
  }\bibfield  {title} {\bibinfo {title} {{Quantum field theory in de Sitter and
  quasi--de Sitter spacetimes revisited}},\ }\href
  {https://doi.org/10.1103/PhysRevD.87.104004} {\bibfield  {journal} {\bibinfo
  {journal} {Phys. Rev. D}\ }\textbf {\bibinfo {volume} {87}},\ \bibinfo
  {pages} {104004} (\bibinfo {year} {2013})},\ \Eprint
  {https://arxiv.org/abs/1302.7177} {arXiv:1302.7177 [gr-qc]} \BibitemShut
  {NoStop}%
\bibitem [{\citenamefont {Steeg}\ and\ \citenamefont
  {Menicucci}(2009)}]{ver2009entangling}%
  \BibitemOpen
  \bibfield  {author} {\bibinfo {author} {\bibfnamefont {G.~V.}\ \bibnamefont
  {Steeg}}\ and\ \bibinfo {author} {\bibfnamefont {N.~C.}\ \bibnamefont
  {Menicucci}},\ }\bibfield  {title} {\bibinfo {title} {Entangling power of an
  expanding universe},\ }\href {https://doi.org/10.1103/PhysRevD.79.044027}
  {\bibfield  {journal} {\bibinfo  {journal} {Phys. Rev. D}\ }\textbf {\bibinfo
  {volume} {79}},\ \bibinfo {pages} {044027} (\bibinfo {year}
  {2009})}\BibitemShut {NoStop}%
\bibitem [{\citenamefont {Nambu}(2013)}]{nambu2013entanglement}%
  \BibitemOpen
  \bibfield  {author} {\bibinfo {author} {\bibfnamefont {Y.}~\bibnamefont
  {Nambu}},\ }\bibfield  {title} {\bibinfo {title} {{Entanglement Structure in
  Expanding Universes}},\ }\href {https://doi.org/10.3390/e15051847} {\bibfield
   {journal} {\bibinfo  {journal} {Entropy}\ }\textbf {\bibinfo {volume}
  {15}},\ \bibinfo {pages} {1847} (\bibinfo {year} {2013})},\ \Eprint
  {https://arxiv.org/abs/1305.4193} {arXiv:1305.4193 [gr-qc]} \BibitemShut
  {NoStop}%
\bibitem [{\citenamefont {Kukita}\ and\ \citenamefont
  {Nambu}(2017)}]{kukita2017entanglement}%
  \BibitemOpen
  \bibfield  {author} {\bibinfo {author} {\bibfnamefont {S.}~\bibnamefont
  {Kukita}}\ and\ \bibinfo {author} {\bibfnamefont {Y.}~\bibnamefont {Nambu}},\
  }\bibfield  {title} {\bibinfo {title} {Entanglement dynamics in de sitter
  spacetime},\ }\href {https://doi.org/10.1088/1361-6382/aa8e31} {\bibfield
  {journal} {\bibinfo  {journal} {Classical and Quantum Gravity}\ }\textbf
  {\bibinfo {volume} {34}},\ \bibinfo {pages} {235010} (\bibinfo {year}
  {2017})}\BibitemShut {NoStop}%
\bibitem [{\citenamefont {Tian}\ and\ \citenamefont
  {Jing}(2014)}]{tian2014dynamics}%
  \BibitemOpen
  \bibfield  {author} {\bibinfo {author} {\bibfnamefont {Z.}~\bibnamefont
  {Tian}}\ and\ \bibinfo {author} {\bibfnamefont {J.}~\bibnamefont {Jing}},\
  }\bibfield  {title} {\bibinfo {title} {Dynamics and quantum entanglement of
  two-level atoms in de sitter spacetime},\ }\href
  {https://doi.org/10.1016/j.aop.2014.07.006} {\bibfield  {journal} {\bibinfo
  {journal} {Annals of Physics}\ }\textbf {\bibinfo {volume} {350}},\ \bibinfo
  {pages} {1–13} (\bibinfo {year} {2014})}\BibitemShut {NoStop}%
\bibitem [{\citenamefont {Tian}\ \emph {et~al.}(2016)\citenamefont {Tian},
  \citenamefont {Wang}, \citenamefont {Jing},\ and\ \citenamefont
  {Dragan}}]{tian2016detecting}%
  \BibitemOpen
  \bibfield  {author} {\bibinfo {author} {\bibfnamefont {Z.}~\bibnamefont
  {Tian}}, \bibinfo {author} {\bibfnamefont {J.}~\bibnamefont {Wang}}, \bibinfo
  {author} {\bibfnamefont {J.}~\bibnamefont {Jing}},\ and\ \bibinfo {author}
  {\bibfnamefont {A.}~\bibnamefont {Dragan}},\ }\bibfield  {title} {\bibinfo
  {title} {Detecting the curvature of de sitter universe with two entangled
  atoms},\ }\bibfield  {journal} {\bibinfo  {journal} {Scientific Reports}\
  }\textbf {\bibinfo {volume} {6}},\ \href {https://doi.org/10.1038/srep35222}
  {10.1038/srep35222} (\bibinfo {year} {2016})\BibitemShut {NoStop}%
\bibitem [{\citenamefont {Salton}\ \emph {et~al.}(2015)\citenamefont {Salton},
  \citenamefont {Mann},\ and\ \citenamefont
  {Menicucci}}]{salton2015acceleration}%
  \BibitemOpen
  \bibfield  {author} {\bibinfo {author} {\bibfnamefont {G.}~\bibnamefont
  {Salton}}, \bibinfo {author} {\bibfnamefont {R.~B.}\ \bibnamefont {Mann}},\
  and\ \bibinfo {author} {\bibfnamefont {N.~C.}\ \bibnamefont {Menicucci}},\
  }\bibfield  {title} {\bibinfo {title} {Acceleration-assisted entanglement
  harvesting and rangefinding},\ }\href
  {https://doi.org/10.1088/1367-2630/17/3/035001} {\bibfield  {journal}
  {\bibinfo  {journal} {New Journal of Physics}\ }\textbf {\bibinfo {volume}
  {17}},\ \bibinfo {pages} {035001} (\bibinfo {year} {2015})}\BibitemShut
  {NoStop}%
\bibitem [{\citenamefont {Mann}\ and\ \citenamefont
  {Ralph}(2012)}]{mann2012relativistic}%
  \BibitemOpen
  \bibfield  {author} {\bibinfo {author} {\bibfnamefont {R.~B.}\ \bibnamefont
  {Mann}}\ and\ \bibinfo {author} {\bibfnamefont {T.~C.}\ \bibnamefont
  {Ralph}},\ }\bibfield  {title} {\bibinfo {title} {Relativistic quantum
  information},\ }\href {https://doi.org/10.1088/0264-9381/29/22/220301}
  {\bibfield  {journal} {\bibinfo  {journal} {Classical and Quantum Gravity}\
  }\textbf {\bibinfo {volume} {29}},\ \bibinfo {pages} {220301} (\bibinfo
  {year} {2012})}\BibitemShut {NoStop}%
\bibitem [{\citenamefont {Oi}(2003)}]{oi2003interference}%
  \BibitemOpen
  \bibfield  {author} {\bibinfo {author} {\bibfnamefont {D.~K.~L.}\
  \bibnamefont {Oi}},\ }\bibfield  {title} {\bibinfo {title} {Interference of
  quantum channels},\ }\href {https://doi.org/10.1103/PhysRevLett.91.067902}
  {\bibfield  {journal} {\bibinfo  {journal} {Phys. Rev. Lett.}\ }\textbf
  {\bibinfo {volume} {91}},\ \bibinfo {pages} {067902} (\bibinfo {year}
  {2003})}\BibitemShut {NoStop}%
\bibitem [{\citenamefont {Chiribella}\ and\ \citenamefont
  {Kristjánsson}(2019)}]{chiribella2019quantum}%
  \BibitemOpen
  \bibfield  {author} {\bibinfo {author} {\bibfnamefont {G.}~\bibnamefont
  {Chiribella}}\ and\ \bibinfo {author} {\bibfnamefont {H.}~\bibnamefont
  {Kristjánsson}},\ }\bibfield  {title} {\bibinfo {title} {Quantum shannon
  theory with superpositions of trajectories},\ }\href
  {https://doi.org/10.1098/rspa.2018.0903} {\bibfield  {journal} {\bibinfo
  {journal} {Proceedings of the Royal Society A: Mathematical, Physical and
  Engineering Sciences}\ }\textbf {\bibinfo {volume} {475}},\ \bibinfo {pages}
  {20180903} (\bibinfo {year} {2019})}\BibitemShut {NoStop}%
\bibitem [{\citenamefont {Kosloff}(2013)}]{kosloffe15062100}%
  \BibitemOpen
  \bibfield  {author} {\bibinfo {author} {\bibfnamefont {R.}~\bibnamefont
  {Kosloff}},\ }\bibfield  {title} {\bibinfo {title} {Quantum thermodynamics: A
  dynamical viewpoint},\ }\href {https://doi.org/10.3390/e15062100} {\bibfield
  {journal} {\bibinfo  {journal} {Entropy}\ }\textbf {\bibinfo {volume} {15}},\
  \bibinfo {pages} {2100} (\bibinfo {year} {2013})}\BibitemShut {NoStop}%
\bibitem [{\citenamefont {Vinjanampathy}\ and\ \citenamefont
  {Anders}(2016)}]{andersdoi:10.1080/00107514.2016.1201896}%
  \BibitemOpen
  \bibfield  {author} {\bibinfo {author} {\bibfnamefont {S.}~\bibnamefont
  {Vinjanampathy}}\ and\ \bibinfo {author} {\bibfnamefont {J.}~\bibnamefont
  {Anders}},\ }\bibfield  {title} {\bibinfo {title} {Quantum thermodynamics},\
  }\href {https://doi.org/10.1080/00107514.2016.1201896} {\bibfield  {journal}
  {\bibinfo  {journal} {Contemporary Physics}\ }\textbf {\bibinfo {volume}
  {57}},\ \bibinfo {pages} {545} (\bibinfo {year} {2016})},\ \Eprint
  {https://arxiv.org/abs/https://doi.org/10.1080/00107514.2016.1201896}
  {https://doi.org/10.1080/00107514.2016.1201896} \BibitemShut {NoStop}%
\bibitem [{\citenamefont {Foo}\ \emph {et~al.}(2020{\natexlab{b}})\citenamefont
  {Foo}, \citenamefont {Onoe},\ and\ \citenamefont
  {Zych}}]{foo2020unruhdewitt}%
  \BibitemOpen
  \bibfield  {author} {\bibinfo {author} {\bibfnamefont {J.}~\bibnamefont
  {Foo}}, \bibinfo {author} {\bibfnamefont {S.}~\bibnamefont {Onoe}},\ and\
  \bibinfo {author} {\bibfnamefont {M.}~\bibnamefont {Zych}},\ }\href@noop {}
  {\bibinfo {title} {Unruh-dewitt detectors in quantum superpositions of
  trajectories}} (\bibinfo {year} {2020}{\natexlab{b}}),\ \Eprint
  {https://arxiv.org/abs/2003.12774} {arXiv:2003.12774 [quant-ph]} \BibitemShut
  {NoStop}%
\bibitem [{\citenamefont {Weldon}(2000)}]{weldon2000thermal}%
  \BibitemOpen
  \bibfield  {author} {\bibinfo {author} {\bibfnamefont {H.~A.}\ \bibnamefont
  {Weldon}},\ }\bibfield  {title} {\bibinfo {title} {Thermal green functions in
  coordinate space for massless particles of any spin},\ }\bibfield  {journal}
  {\bibinfo  {journal} {Physical Review D}\ }\textbf {\bibinfo {volume} {62}},\
  \href {https://doi.org/10.1103/physrevd.62.056010}
  {10.1103/physrevd.62.056010} (\bibinfo {year} {2000})\BibitemShut {NoStop}%
\bibitem [{\citenamefont {Griffiths}\ and\ \citenamefont
  {Podolsk{\`y}}(2009)}]{griffiths2009exact}%
  \BibitemOpen
  \bibfield  {author} {\bibinfo {author} {\bibfnamefont {J.~B.}\ \bibnamefont
  {Griffiths}}\ and\ \bibinfo {author} {\bibfnamefont {J.}~\bibnamefont
  {Podolsk{\`y}}},\ }\href@noop {} {\emph {\bibinfo {title} {Exact space-times
  in Einstein's general relativity}}}\ (\bibinfo  {publisher} {Cambridge
  University Press},\ \bibinfo {year} {2009})\BibitemShut {NoStop}%
\bibitem [{\citenamefont {Kubo}(1957)}]{Kubo:1957mj}%
  \BibitemOpen
  \bibfield  {author} {\bibinfo {author} {\bibfnamefont {R.}~\bibnamefont
  {Kubo}},\ }\bibfield  {title} {\bibinfo {title} {{Statistical mechanical
  theory of irreversible processes. 1. General theory and simple applications
  in magnetic and conduction problems}},\ }\href
  {https://doi.org/10.1143/JPSJ.12.570} {\bibfield  {journal} {\bibinfo
  {journal} {J. Phys. Soc. Jap.}\ }\textbf {\bibinfo {volume} {12}},\ \bibinfo
  {pages} {570} (\bibinfo {year} {1957})}\BibitemShut {NoStop}%
\bibitem [{\citenamefont {Henderson}\ \emph {et~al.}(2020)\citenamefont
  {Henderson}, \citenamefont {Belenchia}, \citenamefont {Castro-Ruiz},
  \citenamefont {Budroni}, \citenamefont {Zych}, \citenamefont {Časlav
  Brukner},\ and\ \citenamefont {Mann}}]{henderson2020quantum}%
  \BibitemOpen
  \bibfield  {author} {\bibinfo {author} {\bibfnamefont {L.~J.}\ \bibnamefont
  {Henderson}}, \bibinfo {author} {\bibfnamefont {A.}~\bibnamefont
  {Belenchia}}, \bibinfo {author} {\bibfnamefont {E.}~\bibnamefont
  {Castro-Ruiz}}, \bibinfo {author} {\bibfnamefont {C.}~\bibnamefont
  {Budroni}}, \bibinfo {author} {\bibfnamefont {M.}~\bibnamefont {Zych}},
  \bibinfo {author} {\bibnamefont {Časlav Brukner}},\ and\ \bibinfo {author}
  {\bibfnamefont {R.~B.}\ \bibnamefont {Mann}},\ }\href@noop {} {\bibinfo
  {title} {Quantum temporal superposition: the case of qft}} (\bibinfo {year}
  {2020}),\ \Eprint {https://arxiv.org/abs/2002.06208} {arXiv:2002.06208
  [quant-ph]} \BibitemShut {NoStop}%
\bibitem [{\citenamefont {Cong}\ \emph {et~al.}(2020)\citenamefont {Cong},
  \citenamefont {Bicak}, \citenamefont {Kubiznak},\ and\ \citenamefont
  {Mann}}]{Cong:2020crf}%
  \BibitemOpen
  \bibfield  {author} {\bibinfo {author} {\bibfnamefont {W.}~\bibnamefont
  {Cong}}, \bibinfo {author} {\bibfnamefont {J.}~\bibnamefont {Bicak}},
  \bibinfo {author} {\bibfnamefont {D.}~\bibnamefont {Kubiznak}},\ and\
  \bibinfo {author} {\bibfnamefont {R.~B.}\ \bibnamefont {Mann}},\ }\bibfield
  {title} {\bibinfo {title} {{Quantum distinction of inertial frames: Local
  versus global}},\ }\href {https://doi.org/10.1103/PhysRevD.101.104060}
  {\bibfield  {journal} {\bibinfo  {journal} {Phys. Rev. D}\ }\textbf {\bibinfo
  {volume} {101}},\ \bibinfo {pages} {104060} (\bibinfo {year} {2020})},\
  \Eprint {https://arxiv.org/abs/2003.09719} {arXiv:2003.09719 [gr-qc]}
  \BibitemShut {NoStop}%
\bibitem [{\citenamefont {Louko}\ and\ \citenamefont
  {Satz}(2008{\natexlab{b}})}]{Louko:2007mu}%
  \BibitemOpen
  \bibfield  {author} {\bibinfo {author} {\bibfnamefont {J.}~\bibnamefont
  {Louko}}\ and\ \bibinfo {author} {\bibfnamefont {A.}~\bibnamefont {Satz}},\
  }\bibfield  {title} {\bibinfo {title} {{Transition rate of the Unruh-DeWitt
  detector in curved spacetime}},\ }\href
  {https://doi.org/10.1088/0264-9381/25/5/055012} {\bibfield  {journal}
  {\bibinfo  {journal} {Class. Quant. Grav.}\ }\textbf {\bibinfo {volume}
  {25}},\ \bibinfo {pages} {055012} (\bibinfo {year} {2008}{\natexlab{b}})},\
  \Eprint {https://arxiv.org/abs/0710.5671} {arXiv:0710.5671 [gr-qc]}
  \BibitemShut {NoStop}%
\bibitem [{\citenamefont {Smith}\ and\ \citenamefont
  {Mann}(2014)}]{Smith:2013zqa}%
  \BibitemOpen
  \bibfield  {author} {\bibinfo {author} {\bibfnamefont {A.~R.~H.}\
  \bibnamefont {Smith}}\ and\ \bibinfo {author} {\bibfnamefont {R.~B.}\
  \bibnamefont {Mann}},\ }\bibfield  {title} {\bibinfo {title} {{Looking Inside
  a Black Hole}},\ }\href {https://doi.org/10.1088/0264-9381/31/8/082001}
  {\bibfield  {journal} {\bibinfo  {journal} {Class. Quant. Grav.}\ }\textbf
  {\bibinfo {volume} {31}},\ \bibinfo {pages} {082001} (\bibinfo {year}
  {2014})},\ \Eprint {https://arxiv.org/abs/1309.4125} {arXiv:1309.4125
  [gr-qc]} \BibitemShut {NoStop}%
\bibitem [{\citenamefont {Langlois}(2006)}]{langlois2006causal}%
  \BibitemOpen
  \bibfield  {author} {\bibinfo {author} {\bibfnamefont {P.}~\bibnamefont
  {Langlois}},\ }\bibfield  {title} {\bibinfo {title} {{Causal particle
  detectors and topology}},\ }\href {https://doi.org/10.1016/j.aop.2006.01.013}
  {\bibfield  {journal} {\bibinfo  {journal} {Annals Phys.}\ }\textbf {\bibinfo
  {volume} {321}},\ \bibinfo {pages} {2027} (\bibinfo {year} {2006})},\ \Eprint
  {https://arxiv.org/abs/gr-qc/0510049} {arXiv:gr-qc/0510049} \BibitemShut
  {NoStop}%
\bibitem [{\citenamefont {Gu\'erin}\ \emph {et~al.}(2016)\citenamefont
  {Gu\'erin}, \citenamefont {Feix}, \citenamefont {Ara\'ujo},\ and\
  \citenamefont {Brukner}}]{guerinPhysRevLett.117.100502}%
  \BibitemOpen
  \bibfield  {author} {\bibinfo {author} {\bibfnamefont {P.~A.}\ \bibnamefont
  {Gu\'erin}}, \bibinfo {author} {\bibfnamefont {A.}~\bibnamefont {Feix}},
  \bibinfo {author} {\bibfnamefont {M.}~\bibnamefont {Ara\'ujo}},\ and\
  \bibinfo {author} {\bibfnamefont {i.~c.~v.}\ \bibnamefont {Brukner}},\
  }\bibfield  {title} {\bibinfo {title} {Exponential communication complexity
  advantage from quantum superposition of the direction of communication},\
  }\href {https://doi.org/10.1103/PhysRevLett.117.100502} {\bibfield  {journal}
  {\bibinfo  {journal} {Phys. Rev. Lett.}\ }\textbf {\bibinfo {volume} {117}},\
  \bibinfo {pages} {100502} (\bibinfo {year} {2016})}\BibitemShut {NoStop}%
\bibitem [{\citenamefont {Chiribella}(2012)}]{chiribellaPhysRevA.86.040301}%
  \BibitemOpen
  \bibfield  {author} {\bibinfo {author} {\bibfnamefont {G.}~\bibnamefont
  {Chiribella}},\ }\bibfield  {title} {\bibinfo {title} {Perfect discrimination
  of no-signalling channels via quantum superposition of causal structures},\
  }\href {https://doi.org/10.1103/PhysRevA.86.040301} {\bibfield  {journal}
  {\bibinfo  {journal} {Phys. Rev. A}\ }\textbf {\bibinfo {volume} {86}},\
  \bibinfo {pages} {040301} (\bibinfo {year} {2012})}\BibitemShut {NoStop}%
\bibitem [{\citenamefont {Abbott}\ \emph {et~al.}(2020)\citenamefont {Abbott},
  \citenamefont {Wechs}, \citenamefont {Horsman}, \citenamefont {Mhalla},\ and\
  \citenamefont {Branciard}}]{Abbott_2020}%
  \BibitemOpen
  \bibfield  {author} {\bibinfo {author} {\bibfnamefont {A.~A.}\ \bibnamefont
  {Abbott}}, \bibinfo {author} {\bibfnamefont {J.}~\bibnamefont {Wechs}},
  \bibinfo {author} {\bibfnamefont {D.}~\bibnamefont {Horsman}}, \bibinfo
  {author} {\bibfnamefont {M.}~\bibnamefont {Mhalla}},\ and\ \bibinfo {author}
  {\bibfnamefont {C.}~\bibnamefont {Branciard}},\ }\bibfield  {title} {\bibinfo
  {title} {Communication through coherent control of quantum channels},\ }\href
  {https://doi.org/10.22331/q-2020-09-24-333} {\bibfield  {journal} {\bibinfo
  {journal} {Quantum}\ }\textbf {\bibinfo {volume} {4}},\ \bibinfo {pages}
  {333} (\bibinfo {year} {2020})}\BibitemShut {NoStop}%
\bibitem [{\citenamefont {Oreshkov}\ \emph {et~al.}(2012)\citenamefont
  {Oreshkov}, \citenamefont {Costa},\ and\ \citenamefont
  {Brukner}}]{oreshkov2012quantum}%
  \BibitemOpen
  \bibfield  {author} {\bibinfo {author} {\bibfnamefont {O.}~\bibnamefont
  {Oreshkov}}, \bibinfo {author} {\bibfnamefont {F.}~\bibnamefont {Costa}},\
  and\ \bibinfo {author} {\bibfnamefont {v.~C.}\ \bibnamefont {Brukner}},\
  }\bibfield  {title} {\bibinfo {title} {Quantum correlations with no causal
  order},\ }\bibfield  {journal} {\bibinfo  {journal} {Nature Communications}\
  }\textbf {\bibinfo {volume} {3}},\ \href {https://doi.org/10.1038/ncomms2076}
  {10.1038/ncomms2076} (\bibinfo {year} {2012})\BibitemShut {NoStop}%
\bibitem [{\citenamefont {Ebler}\ \emph {et~al.}(2018)\citenamefont {Ebler},
  \citenamefont {Salek},\ and\ \citenamefont
  {Chiribella}}]{eblerPhysRevLett.120.120502}%
  \BibitemOpen
  \bibfield  {author} {\bibinfo {author} {\bibfnamefont {D.}~\bibnamefont
  {Ebler}}, \bibinfo {author} {\bibfnamefont {S.}~\bibnamefont {Salek}},\ and\
  \bibinfo {author} {\bibfnamefont {G.}~\bibnamefont {Chiribella}},\ }\bibfield
   {title} {\bibinfo {title} {Enhanced communication with the assistance of
  indefinite causal order},\ }\href
  {https://doi.org/10.1103/PhysRevLett.120.120502} {\bibfield  {journal}
  {\bibinfo  {journal} {Phys. Rev. Lett.}\ }\textbf {\bibinfo {volume} {120}},\
  \bibinfo {pages} {120502} (\bibinfo {year} {2018})}\BibitemShut {NoStop}%
\bibitem [{\citenamefont {Felce}\ and\ \citenamefont
  {Vedral}(2020)}]{felcePhysRevLett.125.070603}%
  \BibitemOpen
  \bibfield  {author} {\bibinfo {author} {\bibfnamefont {D.}~\bibnamefont
  {Felce}}\ and\ \bibinfo {author} {\bibfnamefont {V.}~\bibnamefont {Vedral}},\
  }\bibfield  {title} {\bibinfo {title} {Quantum refrigeration with indefinite
  causal order},\ }\href {https://doi.org/10.1103/PhysRevLett.125.070603}
  {\bibfield  {journal} {\bibinfo  {journal} {Phys. Rev. Lett.}\ }\textbf
  {\bibinfo {volume} {125}},\ \bibinfo {pages} {070603} (\bibinfo {year}
  {2020})}\BibitemShut {NoStop}%
\bibitem [{\citenamefont {Zych}\ \emph {et~al.}(2019)\citenamefont {Zych},
  \citenamefont {Costa}, \citenamefont {Pikovski},\ and\ \citenamefont
  {Brukner}}]{zych2019bell}%
  \BibitemOpen
  \bibfield  {author} {\bibinfo {author} {\bibfnamefont {M.}~\bibnamefont
  {Zych}}, \bibinfo {author} {\bibfnamefont {F.}~\bibnamefont {Costa}},
  \bibinfo {author} {\bibfnamefont {I.}~\bibnamefont {Pikovski}},\ and\
  \bibinfo {author} {\bibfnamefont {v.~C.}\ \bibnamefont {Brukner}},\
  }\bibfield  {title} {\bibinfo {title} {{Bell's theorem for temporal order}},\
  }\href {https://doi.org/10.1038/s41467-019-11579-x} {\bibfield  {journal}
  {\bibinfo  {journal} {Nature Commun.}\ }\textbf {\bibinfo {volume} {10}},\
  \bibinfo {pages} {3772} (\bibinfo {year} {2019})},\ \Eprint
  {https://arxiv.org/abs/1708.00248} {arXiv:1708.00248 [quant-ph]} \BibitemShut
  {NoStop}%
\bibitem [{\citenamefont {Zych}\ \emph {et~al.}(2011)\citenamefont {Zych},
  \citenamefont {Costa}, \citenamefont {Pikovski},\ and\ \citenamefont
  {Brukner}}]{Zych:2011hu}%
  \BibitemOpen
  \bibfield  {author} {\bibinfo {author} {\bibfnamefont {M.}~\bibnamefont
  {Zych}}, \bibinfo {author} {\bibfnamefont {F.}~\bibnamefont {Costa}},
  \bibinfo {author} {\bibfnamefont {I.}~\bibnamefont {Pikovski}},\ and\
  \bibinfo {author} {\bibfnamefont {C.}~\bibnamefont {Brukner}},\ }\bibfield
  {title} {\bibinfo {title} {{Quantum interferometric visibility as a witness
  of general relativistic proper time}},\ }\href
  {https://doi.org/10.1038/ncomms1498} {\bibfield  {journal} {\bibinfo
  {journal} {Nature Commun.}\ }\textbf {\bibinfo {volume} {2}},\ \bibinfo
  {pages} {505} (\bibinfo {year} {2011})},\ \Eprint
  {https://arxiv.org/abs/1105.4531} {arXiv:1105.4531 [quant-ph]} \BibitemShut
  {NoStop}%
\bibitem [{\citenamefont {Zych}\ \emph {et~al.}(2012)\citenamefont {Zych},
  \citenamefont {Costa}, \citenamefont {Pikovski}, \citenamefont {Ralph},\ and\
  \citenamefont {Brukner}}]{Zych:2012ut}%
  \BibitemOpen
  \bibfield  {author} {\bibinfo {author} {\bibfnamefont {M.}~\bibnamefont
  {Zych}}, \bibinfo {author} {\bibfnamefont {F.}~\bibnamefont {Costa}},
  \bibinfo {author} {\bibfnamefont {I.}~\bibnamefont {Pikovski}}, \bibinfo
  {author} {\bibfnamefont {T.~C.}\ \bibnamefont {Ralph}},\ and\ \bibinfo
  {author} {\bibfnamefont {C.}~\bibnamefont {Brukner}},\ }\bibfield  {title}
  {\bibinfo {title} {{General relativistic effects in quantum interference of
  photons}},\ }\href {https://doi.org/10.1088/0264-9381/29/22/224010}
  {\bibfield  {journal} {\bibinfo  {journal} {Class. Quant. Grav.}\ }\textbf
  {\bibinfo {volume} {29}},\ \bibinfo {pages} {224010} (\bibinfo {year}
  {2012})},\ \Eprint {https://arxiv.org/abs/1206.0965} {arXiv:1206.0965
  [quant-ph]} \BibitemShut {NoStop}%
\bibitem [{\citenamefont {Smith}\ and\ \citenamefont
  {Ahmadi}(2020)}]{Smith:2019imm}%
  \BibitemOpen
  \bibfield  {author} {\bibinfo {author} {\bibfnamefont {A.~R.~H.}\
  \bibnamefont {Smith}}\ and\ \bibinfo {author} {\bibfnamefont
  {M.}~\bibnamefont {Ahmadi}},\ }\bibfield  {title} {\bibinfo {title} {{Quantum
  clocks observe classical and quantum time dilation}},\ }\href
  {https://doi.org/10.1038/s41467-020-18264-4} {\bibfield  {journal} {\bibinfo
  {journal} {Nature Commun.}\ }\textbf {\bibinfo {volume} {11}},\ \bibinfo
  {pages} {5360} (\bibinfo {year} {2020})},\ \Eprint
  {https://arxiv.org/abs/1904.12390} {arXiv:1904.12390 [quant-ph]} \BibitemShut
  {NoStop}%
\bibitem [{\citenamefont {Ford}\ and\ \citenamefont
  {Svaiter}(1997)}]{ford1997cosmological}%
  \BibitemOpen
  \bibfield  {author} {\bibinfo {author} {\bibfnamefont {L.~H.}\ \bibnamefont
  {Ford}}\ and\ \bibinfo {author} {\bibfnamefont {N.~F.}\ \bibnamefont
  {Svaiter}},\ }\bibfield  {title} {\bibinfo {title} {Cosmological and black
  hole horizon fluctuations},\ }\href
  {https://doi.org/10.1103/physrevd.56.2226} {\bibfield  {journal} {\bibinfo
  {journal} {Physical Review D}\ }\textbf {\bibinfo {volume} {56}},\ \bibinfo
  {pages} {2226–2235} (\bibinfo {year} {1997})}\BibitemShut {NoStop}%
\bibitem [{\citenamefont {Lake}\ \emph {et~al.}(2019)\citenamefont {Lake},
  \citenamefont {Miller}, \citenamefont {Ganardi}, \citenamefont {Liu},
  \citenamefont {Liang},\ and\ \citenamefont {Paterek}}]{lake2019generalised}%
  \BibitemOpen
  \bibfield  {author} {\bibinfo {author} {\bibfnamefont {M.~J.}\ \bibnamefont
  {Lake}}, \bibinfo {author} {\bibfnamefont {M.}~\bibnamefont {Miller}},
  \bibinfo {author} {\bibfnamefont {R.~F.}\ \bibnamefont {Ganardi}}, \bibinfo
  {author} {\bibfnamefont {Z.}~\bibnamefont {Liu}}, \bibinfo {author}
  {\bibfnamefont {S.-D.}\ \bibnamefont {Liang}},\ and\ \bibinfo {author}
  {\bibfnamefont {T.}~\bibnamefont {Paterek}},\ }\bibfield  {title} {\bibinfo
  {title} {{Generalised uncertainty relations from superpositions of
  geometries}},\ }\href {https://doi.org/10.1088/1361-6382/ab2160} {\bibfield
  {journal} {\bibinfo  {journal} {Class. Quant. Grav.}\ }\textbf {\bibinfo
  {volume} {36}},\ \bibinfo {pages} {155012} (\bibinfo {year} {2019})},\
  \Eprint {https://arxiv.org/abs/1812.10045} {arXiv:1812.10045 [quant-ph]}
  \BibitemShut {NoStop}%
\bibitem [{\citenamefont {Bose}\ \emph {et~al.}(2017)\citenamefont {Bose},
  \citenamefont {Mazumdar}, \citenamefont {Morley}, \citenamefont {Ulbricht},
  \citenamefont {Toro\ifmmode~\check{s}\else \v{s}\fi{}}, \citenamefont
  {Paternostro}, \citenamefont {Geraci}, \citenamefont {Barker}, \citenamefont
  {Kim},\ and\ \citenamefont {Milburn}}]{bose2017spin}%
  \BibitemOpen
  \bibfield  {author} {\bibinfo {author} {\bibfnamefont {S.}~\bibnamefont
  {Bose}}, \bibinfo {author} {\bibfnamefont {A.}~\bibnamefont {Mazumdar}},
  \bibinfo {author} {\bibfnamefont {G.~W.}\ \bibnamefont {Morley}}, \bibinfo
  {author} {\bibfnamefont {H.}~\bibnamefont {Ulbricht}}, \bibinfo {author}
  {\bibfnamefont {M.}~\bibnamefont {Toro\ifmmode~\check{s}\else \v{s}\fi{}}},
  \bibinfo {author} {\bibfnamefont {M.}~\bibnamefont {Paternostro}}, \bibinfo
  {author} {\bibfnamefont {A.~A.}\ \bibnamefont {Geraci}}, \bibinfo {author}
  {\bibfnamefont {P.~F.}\ \bibnamefont {Barker}}, \bibinfo {author}
  {\bibfnamefont {M.~S.}\ \bibnamefont {Kim}},\ and\ \bibinfo {author}
  {\bibfnamefont {G.}~\bibnamefont {Milburn}},\ }\bibfield  {title} {\bibinfo
  {title} {Spin entanglement witness for quantum gravity},\ }\href
  {https://doi.org/10.1103/PhysRevLett.119.240401} {\bibfield  {journal}
  {\bibinfo  {journal} {Phys. Rev. Lett.}\ }\textbf {\bibinfo {volume} {119}},\
  \bibinfo {pages} {240401} (\bibinfo {year} {2017})}\BibitemShut {NoStop}%
\bibitem [{\citenamefont {Christodoulou}\ and\ \citenamefont
  {Rovelli}(2019)}]{christodoulou2019possibility}%
  \BibitemOpen
  \bibfield  {author} {\bibinfo {author} {\bibfnamefont {M.}~\bibnamefont
  {Christodoulou}}\ and\ \bibinfo {author} {\bibfnamefont {C.}~\bibnamefont
  {Rovelli}},\ }\bibfield  {title} {\bibinfo {title} {On the possibility of
  laboratory evidence for quantum superposition of geometries},\ }\href
  {https://doi.org/10.1016/j.physletb.2019.03.015} {\bibfield  {journal}
  {\bibinfo  {journal} {Physics Letters B}\ }\textbf {\bibinfo {volume}
  {792}},\ \bibinfo {pages} {64–68} (\bibinfo {year} {2019})}\BibitemShut
  {NoStop}%
\bibitem [{\citenamefont {Marletto}\ and\ \citenamefont
  {Vedral}(2017)}]{marletto2017gravitationally}%
  \BibitemOpen
  \bibfield  {author} {\bibinfo {author} {\bibfnamefont {C.}~\bibnamefont
  {Marletto}}\ and\ \bibinfo {author} {\bibfnamefont {V.}~\bibnamefont
  {Vedral}},\ }\bibfield  {title} {\bibinfo {title} {Gravitationally induced
  entanglement between two massive particles is sufficient evidence of quantum
  effects in gravity},\ }\href {https://doi.org/10.1103/PhysRevLett.119.240402}
  {\bibfield  {journal} {\bibinfo  {journal} {Phys. Rev. Lett.}\ }\textbf
  {\bibinfo {volume} {119}},\ \bibinfo {pages} {240402} (\bibinfo {year}
  {2017})}\BibitemShut {NoStop}%
\end{thebibliography}%

\end{document}